\documentclass[11pt,twoside]{article}
\usepackage{geometry}
\usepackage{moreverb,url}
\usepackage[normalem]{ulem}   % for use of \sout
\usepackage{graphicx}
\usepackage[colorlinks,bookmarksopen,bookmarksnumbered,citecolor=red,urlcolor=red]{hyperref}
\usepackage{booktabs}
\usepackage{multirow}
\usepackage{footnote}
\usepackage{float}
\usepackage{multirow}
\usepackage{rotating}

\makesavenoteenv{tabular}

\newcommand\BibTeX{{\rmfamily B\kern-.05em \textsc{i\kern-.025em b}\kern-.08em
T\kern-.1667em\lower.7ex\hbox{E}\kern-.125emX}}

\newcommand{\ndz}{{\color{white}0}}

\begin{document}

%\runninghead{Comparison of adverse event risks}

\title{\Large Survival analysis for AdVerse events with VarYing follow-up times (SAVVY) -- comparison of adverse event risks in randomized controlled trials}

\author{Kaspar Rufibach$^{1,*}$, Regina Stegherr$^{2,*}$, Claudia Schmoor$^{3}$, Valentine Jehl$^{4}$, \\Arthur Allignol$^{5}$, Annette Boeckenhoff$^{6}$, Cornelia Dunger-Baldauf$^{4}$, Lewin Eisele$^{7}$,\\ Thomas K\"unzel$^{1}$, Katrin Kupas$^{8}$, Friedhelm Leverkus$^{9}$, Matthias Trampisch$^{10}$, Yumin Zhao$^{11}$, \\Tim Friede$^{12}$and Jan Beyersmann$^{2,**}$ }
\date{September 20, 2022}
\maketitle
\noindent${}^{1}$ F. Hoffmann-La Roche, Basel, Switzerland\\
\noindent${}^{2}$ Institute of Statistics, Ulm University, Ulm, Germany\\
\noindent${}^{3}$ Clinical Trials Unit, Faculty of Medicine and Medical Center, University of Freiburg, Freiburg im Breisgau, Germany\\
\noindent${}^{4}$ Novartis Pharma AG, Novartis Pharma AG, Basel, Switzerland\\
\noindent${}^{5}$ Merck KGaA, Darmstadt, Germany\\
\noindent${}^{6}$ Bayer AG, Wuppertal, Germany\\
\noindent${}^{7}$ Janssen-Cilag GmbH, Neuss, Germany\\
\noindent${}^{8}$ Bristol-Myers-Squibb GmbH \& Co. KGaA, München, Germany\\
\noindent${}^{9}$ Pfizer, Berlin, Germany\\
\noindent${}^{10}$ Boehringer Ingelheim Pharma GmbH \& Co. KG, Ingelheim, Germany\\
\noindent${}^{11}$ Eli Lilly and Company, Indianapolis, Indiana, USA\\
\noindent${}^{12}$ Department of Medical Statistics, University Medical Center G\"ottingen, G\"ottingen, Germany\\
$^*$: Kaspar Rufibach and Regina Stegherr contributed equally to this work

\noindent${}^{**}$ {Corresponding author: Jan Beyersmann, jan.beyersmann@uni-ulm.de}

\begin{abstract}\noindent
%\begin{abstract}
Analyses of adverse events (AEs) are an important aspect
of the evaluation of experimental therapies. The SAVVY (Survival analysis
for AdVerse events with Varying follow-up times) project aims to improve the
analyses of AE data in clinical trials through the use of survival
techniques appropriately dealing with varying follow-up times, censoring,
and competing events (CE). In an empirical study including seventeen
randomized clinical trials the effect of varying follow-up times, censoring,
and competing events on comparisons of two treatment arms with respect to AE risks is investigated.

The comparisons of relative risks (RR) of standard probability-based
estimators to the gold-standard Aalen-Johansen estimator or hazard-based
estimators to an estimated hazard ratio (HR) from Cox regression are done
descriptively, with graphical displays, and using a random effects
meta-analysis on AE level. The influence of different factors on the size of
the bias is investigated in a meta-regression.

We find that for both, avoiding bias and categorization of evidence with
respect to treatment effect on AE risk into categories, the choice of the
estimator is key and more important than features of the underlying data
such as percentage of censoring, CEs, amount of follow-up, or value of the
gold-standard RR.

There is an urgent need to improve the guidelines of reporting AEs so that
incidence proportions are finally replaced by the Aalen-Johansen
estimator --- rather than by Kaplan-Meier --- with appropriate
definition of CEs. For RRs based on hazards, the HR based on Cox regression
has better properties than the ratio of incidence densities.
\end{abstract}


\maketitle

\footnotetext{\textbf{Abbreviations:} AE, adverse event; CE, competing event; HR, hazard ratio; SAVVY, Survival analysis for AdVerse Events with VarYing follow-up times; RR, relative risk}

% -----------------------------------------------------------------------------
\section{Introduction}
%-----------------------------------------------------------------------------

% \kr{Tim/Jan: I had started the "Background" section by copy-and-pasting from 1-sample and then adapt. You asked for marking these passages and offered to shorten. Pls go ahead!}

% \kr{Tim: Asked to summarize most important points from 1-sample paper here in
%   background.}

Methods commonly employed in randomized clinical trials (RCT) to quantify absolute adverse event (AE) risk either do not account for
  varying follow-up times, censoring, or for competing events (CE), although appreciation of these are important in risk quantification, see e.g. Proctor and Schumacher \cite{proctor_16}. % and Putter et al.\cite{putter_07}.
  As noted in Koller et al. \cite{koller_12}%, although large developments in competing events methodology have been achieved over the last decades,
  recognition of competing events in the clinical trial community remains marginal.
  %Methods not accounting for the above features may be grouped into methods that either under- or overestimate cumulative AE probabilities in a time-to-first-event context.

 The SAVVY project group is a collaborative effort from academia and pharmaceutical industry
  with the aim to improve analyses of AE data in clinical trials through use of survival techniques that account for varying follow-up
  times, censoring and CEs. To this end, Stegherr et al. \cite{onesample}, considering only one
  trial arm in an opportunistic set of 186 types of AEs from seventeen RCTs from different disease indications, have
  verified the relevance of using the Aalen-Johansen
  estimator \cite{allignol_16} as the non-parametric gold-standard method when
  quantifying absolute AE risk. The reason is that the Aalen-Johansen
  estimator is the only (non-parametric) estimator that accounts for CEs, censoring, and varying follow-up times simultaneously, and,
  being non-parametric, does not rely on restrictive parametric assumptions such as incidence densities do.

Stegherr et al. confirmed that % Time-to-event or survival endpoints are common in clinical research\cite{Hort:Switz:2005,sato2017statistical}. The observation of the event times is typically incomplete as a consequence of censoring, and the statistical analysis therefore requires specialized techniques. This requirement holds for both the evaluation of efficacy and of safety. An important aim of the latter is the estimation of the probability of an adverse event (AE) of a specific type, which, in a time-to-first-event analysis, is often done by the incidence proportion, i.e., the number of patients with an observed AE (of a certain type) divided by arm size, or the (exposure adjusted) incidence density, which divides by cumulative patient-time at risk. The worry is that the incidence proportion underestimates the cumulative AE probability because it does not account for censoring\cite{oneill_1987,allignol_16,bender_16,unkel_19}. One minus Kaplan-Meier estimator counting AEs as the event would account for censoring, but not for CEs such as death without prior AE. When CEs are present, Kaplan-Meier is commonly used\cite{schumacher2016competing,van2016competing} but bound to overestimate the cumulative AE probability as the methodology implicitly assumes that every
% patient experiences the AE under consideration, possibly after study closure. The incidence density also accounts for censoring, but does not estimate a probability. Rather, it estimates the AE hazard assuming it to be time-constant, and the worry is that this assumption is too restrictive\cite{Krae:even:2009, bender_19}.
%
% The SAVVY project group (Survival analysis for AdVerse events with Varying follow-up times) is a collaborative effort from academia and industry with the aim to improve the analyses of AE data in clinical trials through the use of survival techniques that account for both varying follow-up times, censoring and CEs. In a companion paper\cite{onesample}, the amount of bias when quantifying absolute AE risk in one-sample is assessed when comparing the above estimators to the non-parametric gold-standard, the Aalen-Johansen estimator\cite{allignol_16}. This, in an empirical study on an opportunistic set of seventeen randomized clinical trials (RCT) from different disease indications. In addition, the impact on categorization into AE frequency categories is analyzed.
%
% \kr{TF: We could simply refer here to the one-sample paper and possibly summarize the main results}
the one minus Kaplan-Meier estimator with simple censoring of CEs
overestimates the cumulative AE probability. This is well-known and has been shown
either empirically in one single study \cite{schuster_20} or analytically \cite{stegherr_20}.
Here we extend these considerations to comparing AE risk between treatment arms, the challenge being that, say, the fact of overestimation in both arms of the same trial allows for both under- and overestimation of RRs when comparing arms. We consider the same six estimators of AE probabilities as
in Stegherr et al. \cite{onesample} and extend the results
%on the bias of estimation of {\it absolute} AE probabilities in one sample when compared to the gold-standard Aalen-Johansen estimator
to an assessment of how these same estimators perform when estimating {\it relative} risks between two randomized treatment arms. With this, we answer a question
raised in Unkel et al. \cite{unkel_19}, namely which direction the bias goes
for an estimator of a RR when based on biased one-sample
estimators. Since in applications very often the RR for AEs
is not (only) quantified via estimates of AE probabilities but also using an
estimate of the hazard ratio (HR) of AE hazards, we extend the
analysis to two hazard-based estimators of RR. Here, on the one
  hand, we investigate to which extent incidence densities may be used to
  approximate HR estimates from a semi-parametric Cox model. On the
  other hand, we investigate the relevance of CEs and their CE-specific hazards by investigating conclusions either based on a Cox
  model for time to AEs or based on a ratio of probabilities using the Aalen-Johansen estimator. Properties
and estimands of various estimators to quantify RRs between two randomized
arms, and when to prefer which, are discussed elsewhere \cite{unkel_19, stegherr_19}.

The primary objective of the present paper is to provide an empirical
quantification of how large the bias for the estimation of relative AE risk
compared to the gold-standard Aalen-Johansen estimator can become when using
other estimators, in the presence of one or more of varying follow-up times,
censoring, or CEs. To this end, we accept common choices of
  safety patient data sets which may differ from the intention-to-treat data
  sets. Motivated by the common use of incidence proportions, we assume that
  primary interest is in probabilities, but that estimation of AE
  probabilities must account for varying follow-up times and censoring as
  illustrated in discussions on contrasting incidence proportions and incidence
  densities.

% -----------------------------------------------------------------------------
\section{Methods}
% -----------------------------------------------------------------------------

In the entire paper, Arm E is referring to the experimental treatment and Arm C to the control. %Follow-up is quantified using the median of all event times, whether censored or not.

{\color{black}
% -----------------------------------------------------------------------------
\subsection{Target of estimation}
% -----------------------------------------------------------------------------

As a consequence of our focus on probabilities, the target of estimation or estimand is $P(\mbox{AE in\ }[0,t])$ to be compared between arms. All the five estimation methods that we compare in what follows target this same estimand. In situations not additionally complicated by varying follow-up times or censoring, i.e. when all patients are observed for the same amount of time, $P(\mbox{AE in\ }[0,t])$ can easily be estimated using the incidence proportion. However, as soon as we have varying follow-up and/or censoring, the incidence proportion will typically be a biased estimate of $P(\mbox{AE in\ }[0,t])$.

While we are not at all attempting to define what a fit-for-purpose estimand to quantify safety risk could be, but rather focus on statistical properties of commonly used estimators in the presence of varying follow-up and CEs, a reviewer encouraged us to formulate the five attributes of our target of estimation within the ICH E9(R1) estimand framework. {\it Treatment} and {\it population} are generic, the {\it summary measure} is the relative risk based on arm-wise probabilities $P(\mbox{AE\ in\ }[0,t])$. The {\it variable} should be interpreted in a stochastic process formulation (the backbone of survival analysis), where it is time to composite, i.e. {\it time-to-first-of \{AE, competing event\}}, with indication of type of event (a "mark" in stochastic process language). This implies for {\it intercurrent events} (ICE) that competing events do not affect the \emph{existence of the measurements}, because the different competing events are simply different values of precisely one random variable. One could argue that competing events are thus simply made part of the {\it variable} attribute of the estimand.
}

% -----------------------------------------------------------------------------
\subsection{Organisation of data analysis}
% -----------------------------------------------------------------------------
 In a big collaborative effort data from seventeen RCTs from different disease indications has been gathered within ten sponsor organisations (nine pharmaceutical companies and one academic trial center). In order to avoid challenges with data sharing SAVVY used an approach familiar from Health Informatics, see e.g. Budin et al. \cite{budin_15}. A standardized data structure was defined \cite{stegherr_19} based on which SAS and R macros were developed by the academic project group members. These macros where then shared with all participating sponsor organisations and run by them locally on their individual trial data. Only aggregated data necessary for meta-analyses were forwarded to the academic group members to centrally run meta-analyses.

% -----------------------------------------------------------------------------
\subsection{Focus on AEs}
% -----------------------------------------------------------------------------
In SAVVY we focus on analysis of AEs, as opposed to efficacy endpoints as in
Austin and Fine \cite{austin_17a}. This has several reasons: First,
acknowledging that estimates of AE risks based on the incidence proportion are
very prevalent in RCT reporting and may have major impact on labeling and use
of drugs, we consider it paramount to provide good estimates of such
risks. Illustrating the biases empirically we aim to change the default
methods for AE analysis in the future. Second, for AEs we can use generic CEs
such as death or discontinuation of treatment{\color{black}, the former
  preventing an AE from happening, the latter changing its probability of
  occurrence \cite{Gool:Leis:Crow:Stor:esti:1999}}. Again, looking at primary
endpoint data most likely would have required a case-by-case consideration of
what constitutes a CE. And thirdly, given that our primary objective is an
empirical quantification of the bias of often-used estimators, it is useful to
have a substantial amount of data. We are analyzing 186 types of AEs,
collecting the same number of time-to-event primary endpoints (including
definition of the relevant CEs) would most likely have been impossible.

% -----------------------------------------------------------------------------
\subsection{Follow-up time}
% -----------------------------------------------------------------------------

Since for a time-to-event endpoint, both AE probabilities and the amount of
censoring \cite{pocock_02} are time-dependent, we will consider different
evaluation times called~$\tau$. These evaluation times either imposed no
restriction, i.e., evaluated the estimators until the maximum follow-up time
($\tau_E$ and $\tau_C$ in each arm, respectively), or considered the minimum
of quantiles of observed times in the two treatment arms; the quantiles were
100\% (whose minimum over both arms we denote by
$\tau_{max} = \min\{\tau_E, \tau_C\}$), 90\%, 60\% and 30\%. For computation
of RRs by the ratio of AE probabilities, in this paper we always use the same
follow-up time in both arms, i.e. $\tau_{max}$. The rationale to evaluate
estimators at the latest meaningful time is that this reflects common practice
for the estimation of AE risk using the incidence proportion, where simply the
number of patients with a certain AE in a given time interval is divided
through the total number of patients. Analyses for maximum follow-up (which
may be arm-specific) were not performed. For hazard-based estimators we use
all available data and exclusively look at the arm-specific maximum follow-up
time, i.e. $\tau_E$ and $\tau_C$, also when comparing arms. The partial
likelihood estimator of the HR based on Cox regression ``self-adjusts'' for
$\tau_{max}$ in that it only requires data up to this time point. To limit the
variability of the Cox HR we limit the considered dataset to those AEs with a{\color{black}n absolute}
frequency of $\ge 10$ in {\it each arm} for hazard-based
analyses{\color{black}; we refer to the book by Therneau and Grambsch
  \cite{Terry} for guidance on Cox regression with rare events.}

% \kr{Jan, zu deiner Frage: Fuer hazard-based nehmen wir alle
%   verfuegbaren Daten in jedem Arm, wogegen wir fuer prob-based jeweils nur
%   Daten bis $\tau_{max}$ betrachten. Falls du das nicht klar genug findest -
%   koenntest du das anpassen?}

\subsection{Estimators of AE probabilities}

As one-sample estimators of the AE probability we use the incidence
proportion, the probability transform incidence density ignoring and
accounting for CE, one minus Kaplan-Meier, and the Aalen-Johansen estimator. A
brief summary of one-sample estimators is given in Stegherr et
al. \cite{onesample}. An even more detailed Statistical Analysis Plan for the
entire SAVVY project has been published elsewhere
\cite{stegherr_19}. {\color{black}The probability transforms of the incidence
  density are parametric counterparts based on the exponential distribution of one minus Kaplan-Meier (ignoring CEs)
  and of Aalen-Johansen (accounting for CEs), respectively. Transforms are
  required because incidence densities estimate hazards.}

\subsection{Hazard-based analyses}

For a time-to-event endpoint, the incidence density estimates the hazard function assuming that it is constant \cite{stegherr_19, onesample}. Their ratio therefore estimates a HR under the restrictive assumption that the hazard functions of both treatment arms are constant. The common Cox model is a semi-parametric extension which only requires the HR to be constant, but not the hazards in the arms.

If we consider a time-to-event endpoint with just one possible event, e.g. death for overall survival, then one minus the survival function, i.e. the expected proportion of patients with the event of interest over time, bears a direct relationship to the (cumulative) hazard. However, once we have to consider CEs, the direct relation between the hazard for a given event and the cumulative incidence function, which takes the role of one minus the survival (or ``event-probability'') function, breaks down \cite{latouche_13}. As a consequence, if we want to use a hazard-based analysis to quantify the effect of treatment on the event of interest and all CEs, in theory it is necessary to report {\it all} event-specific hazards, or rather the corresponding HRs. For this reason, we will not only report the performance of hazard-based RR estimators for the endpoint of interest, time to (first) AE, but also for the competing endpoints time to a CE of death and time to all CEs (see ``Definition of CEs'' in Stegherr et al. \cite{onesample} and section below). This is in perfect analogy to not only consider a (one minus Kaplan-Meier-like) simple probability transformation of the AE incidence density, but to also consider an (Aalen-Johansen-like) transformation accounting for CEs.

\subsection{Effect measures}

For given estimators $\hat q_E$ and $\hat q_C$ of AE probabilities calculated at a specific evaluation time within each treatment arm, one can consider either
the risk difference, $\widehat{RD} = \hat q_E - \hat q_C$, the RR,
$\widehat{RR} = \hat q_E / \hat q_C$, or the odds ratio,
$\widehat{OR} = \hat q_E / (1 - \hat q_E)/ (\hat q_C / (1 - \hat q_C))$. In
this paper, we will focus on the RR to quantify the treatment effect. The IQWiG
summarizes reasons to prefer the RR over the risk difference in Appendix
A \cite{iqwigmethodenpaper} of their general methods. The key feature therein that leads us to prefer
the RR is that the risk difference is an absolute effect measure and as such
strongly depends on the baseline risk in the control arm. % Furthermore, variance
% estimators are more easily obtained for RR than risk differences, because
% ratios are amenable to application of the delta rule and no bootstrapping to
% receive variances of RRs is necessary.\kr{Jan: Ich bilde mir ein, diese
%   Erklaerung von Regina so erhalten zu haben? Koenntet ihr euch allenfalls
%   rasch kurzschliessen dazu?}
Finally, we prefer the RR over the OR because it is easier to interpret in the sense that it is an immediate comparison of the cumulative
  AE probabilities estimated in the one-sample case \cite{onesample}.  % \kr{Tim is asking:
  % "Should we also include a comment on OR?"  What would be the key argument
  % against using OR?  I have given a pretty basic and debatable one, are there
  % more?}
%Noch eine genuere Erklaerung dazu: Die Varianz des (log-)ratio der RRs kann man aus den Varianzen der (log-) AE probabilities mittels delta-methode berechnen, da ja z.b. %(IP_E/IP_C)/(AJE_E/AJE_C)=(IP_E/AJE_E)/(IP_C/AJE_C) und daher ein schoener Zusammenhang besteht. Also wurde die Varianz des log-ratio der RR nicht extra gebootstrapped, sondern aus des Varianzen der log-ratios der AE %probs in beiden Gruppen berechnet. Das Ratio der RD ist ja (IP_E-IP_C)/(AJE_E-AJE_C). Ich sehe zumindest nicht, dass ich das aus IP_E/AJE_E und IP_C/AJE_C berechnen kann.
Variance estimators are easily obtained via the delta rule. In the one-sample case, estimates of AE probabilities were benchmarked on the
gold-standard Aalen-Johansen estimator with the primary definition of CEs, i.e. considering all clinical events described below as CEs. This, because the latter is a fully
nonparametric estimator that accounts for censoring, does not rely on a
constant hazard assumption, and accounts for CEs, see Table~1 in Stegherr et al. \cite{onesample}. Furthermore, as is
well-known, simply taking one minus Kaplan-Meier for time-to-first-AE is a
biased estimator of the AE probability in presence of CEs. Here,
as a straightforward extension for the comparison of AE probabilities between
two arms using the RR, we benchmark the latter on the RR estimated using the
Aalen-Johansen estimator in each arm.

The gold-standard for estimates of the HR will be the HR from Cox
regression. This, because the latter is typically used to quantify a treatment
effect not only for efficacy, but also for time-to-first-AE type
endpoints. Variances of comparisons of different estimators of the HR will be
received via bootstrapping \cite{stegherr_20}. The reason to bootstrap is that
we compare different estimators of the same quantity based on one data set,
leading to the different estimators being dependent. {\color{black}Note that
  Cox regression for HRs of an AE technically censors other CEs. This is
  justified because it does not alter the signal of the AE-specific counting
  process, ensuring identifiability of AE hazards and their ratios. The link
  to our earlier probability considerations is that probability transforms
  must account for the presence of multiple hazards (and HRs).}
A Kaplan-Meier estimator that only counts AEs and censors other CEs therefore has no proper probability information. It represents some transformation
  from the hazard scale, and comparing such Kaplan-Meier estimators between
  treatment groups may be used to illustrate treatment contrasts on the AE
  hazard, but such Kaplan-Meier estimators do not have a proper probability
  interpretation as consequence of multiple hazards being present.

As we base our analyses on real datasets we do not know the underlying true effects, either RRs or HRs. This is why we chose the above gold-standard estimators to benchmark the candidate estimators against. In what follows, we will still call the deviation of an estimator under consideration to its respective gold-standard {\it bias}, although, of course, the comparison is between estimators. We note however that, say, a comparison of one minus Kaplan-Meier and Aalen-Johansen will converge in probability towards the true or asymptotic bias when sample size tends to infinity. As discussed above, the reason not to benchmark against the true value in a simulation setup is that in SAVVY we are explicitly interested in quantifying {\it empirically} how much a given estimator can differ compared to the gold-standard Aalen-Johansen estimator.

\subsection{Definition of CEs}
\label{def:CE}

%\kr{CDB: some comments about the CE definition: (1) It is different from Allignol 2016 and Gooley (1999). These refer to events which alter the probability of or preclude the occurrence of the event of interest. Here the recording is an additional criterion. It would be useful to comment. on the reason (2) Was it implemented this way by the companies contributing studies? (3) Not clear to me how the setting of no censoring implies the CE definition here.}

The definition of events as CE (or ``competing risk'') is discussed in detail
in Stegherr et al. \cite{onesample}. Briefly, both death before AE and any
event that would \emph{both} be viewed from a patient
perspective as an event of his/her course of disease or treatment
\emph{and} would stop the recording of the AE of
interest will be viewed as a CE, including possibly disease- or safety-related
loss to follow-up, withdrawal of consent, and
discontinuation. {\color{black}Note that stop of AE recording is included here
  as a CE in situations where the probability of AE occurrence will have
  changed. One example is progression of disease that
    leads to treatment discontinuation and subsequent end of AE recording.

  While this is} our primary definition of a CE{\color{black},} we
will also look at a CE of ``death only''. Even though interpretation depends
on the severity of the event, a categorization into these two types of CEs is
considered here for illustrative purposes. One motivation is that, in a
time-to-first-event analysis, the incidence proportion, that is, the number of
AEs divided by arm size, should be unbiased in the absence of censoring. To
investigate the impact of our primary CE definition, we also included an
investigation of an estimator \emph{Aalen-Johansen (death only)}, which only
treated death before AE as competing, but not the other CEs that belong to our
primary definition.

{\color{black} As a reviewer noted, consideration of ``death only'' reduces the list of ICEs and thus changes the estimand. One can argue that we implicitly define a "while on observation" estimand. This, because even we do not consider the other CEs as ICEs anymore, data collection has still stopped once they have occurred. However, as discussed above, changing the estimand by considering a different set of CEs still begs the question how the different estimation methods perform comparatively. }

% The definition of events as `competing' is essential to both the Aalen-Johansen estimator and the competing incidence density. CEs (or `competing risks') are events that preclude the occurrence of the AE under consideration in a time-to-first-event analysis. One important CE is death before AE. In addition, any event that would both be viewed from a patient perspective as an event of his/her course of disease or treatment and would stop the recording of the interesting AE will be viewed as a CE. To illustrate, discontinuation of trial treatment which mostly leads to end of AE recording will be handled as a CE\cite{jbcs2019}. Consequently, possibly disease- or safety-related loss to follow-up, withdrawal of consent and discontinuation is handled as a CE.

% In order to investigate the impact of the definition of CEs, we also investigated a `death only' scenario, which only treated death before AE as competing, but not the other CEs. This estimator will be called \emph{Aalen-Johansen (death only)} in the following.

An overview how the different estimators account for the three sources of bias, i.e., censoring, no constant hazards, and CEs is given in Table~1 in Stegherr et al. \cite{onesample}.

%%\kr{I moved the following to here, from the section "Role of censoring". I hope that makes sense.}
To explicitly investigate the role of censoring without the methodological complication of CEs, a composite endpoint where AEs and CEs are combined into one single event is considered. As a consequence, the gold-standard in this setting is the one minus Kaplan-Meier estimator which is compared to the incidence proportion.

%\begin{table*}
%\small\sf\centering
%\begin{tabular}{lccc}
% \toprule
%                                                           & Accounts for & Makes no constant     & Accounts for     \\
%                                                           & censoring    & hazard assumption     & CEs              \\
%\midrule
%Incidence proportion                                       & No           & Yes                   & Yes              \\
%Probability transform incidence density ignoring CEs       & Yes          & No (AE Hazard)        & No               \\
%One minus Kaplan-Meier                                     & Yes          & Yes                   & No               \\
%Probability transform incidence density accounting for CEs & Yes          & No (AE and CE Hazard) & Yes              \\
%Aalen-Johansen (death only)                                & Yes          & Yes                   & Yes (Death only) \\
%gold-standard Aalen-Johansen                               & Yes          & Yes                   & Yes              \\
%\bottomrule
%\end{tabular}
%\caption{{\color{black}Overview if the estimators deal with the possible sources of bias.}
%\label{tab:14top}}
%\end{table*}

\subsection{Random effects meta-analysis and meta-regression}
In the meta-analysis and meta-regression, the ratios of the RR estimates, either based on probability or hazard estimates, obtained with one of the other estimators divided by the RR estimate obtained with the gold-standard Aalen-Johansen estimator (for probabilities) or the Cox regression HR (for hazard based) are considered on the log-scale. The standard errors of these log-ratios are calculated with a bootstrap. Then, a normal-normal hierarchical model is fitted and the exponential of the resulting estimate can be interpreted as the average ratio of the two RR estimators.

In a meta-regression it is further investigated which variables impact this average ratio. Therefore, the proportion of censoring, the proportion of CEs, the evaluation time point $\tau$ in years, and the size of the RR under consideration estimated by the gold-standard are included as covariates in an univariable and a multivariable meta-regression. For the latter, since the sum of the proportion of censoring, the proportion of CEs, and the value of the gold-standard Aalen-Johansen estimator converge to 1, including all of them in the model would lead to collinearity. For that reason, we omit the proportion of CEs in the model. All covariates are centered in the meta-regressions.

\subsection{RR categories}
%%\kr{I have updated this, to reflect the earlier discussion of the "category" section below.}

The impact of the use of the different estimators on the conclusions derived from the comparison of treatment arms is investigated by the use of categories. These are typically derived from comparing the confidence interval (CI) of the RR to thresholds. There is no universally accepted standard how one should combine a point estimate and its associated variability, in our case RR, into evidence categories. As an example, we use a categorization motivated by the methods put forward by the IQWiG \cite{iqwigmethodenpaper} for severe AEs (Table 14) to be used for the German benefit-risk assessment. In contrast to the usual IQWiG procedure, however, we do not only categorize the benefit of a therapy, but also the harm. Thereby, in this first analysis we do not distinguish between a positive and a negative treatment effect. Four categories are possible: (0) ``no effect'' if 1 is included in the CI, (a) ``minor'' (``gering'') if the upper bound of the CI is in the interval $[0.9, 1)$ for a RR$<1$ or the lower bound in the interval $(1,1.11]$ for a RR$>1$, (b) ``considerable'' (``betr\"achtlich'') if the upper bound of the CI is in the interval $[0.75,0.9)$ for a RR$<1$ or the lower bound in the interval $(1.11,1.33]$ for a RR$>1$, and (c) ``major'' (``erheblich'') if the upper bound is smaller than 0.75 for a RR$<1$ or the lower bound greater than 1.33 for a RR$>1$. The same categorization is used for the HR instead of RR.

%\begin{figure*}[ht]
%\centering
%\includegraphics[scale=0.7]{plots/fig10_conclusions.png}
%\caption{Categories of the conclusions that can be drawn from the CI of the RR. RR on the $y$-axis. The horizontal reference lines indicate the cutoffs used by IQWiG to categorize effect sizes. For illustration only, no real data displayed.}
%\label{fig10:iqwig}
%%{\color{green}Regina: Can you please add $x$-axis (e.g. "Effect size") and $y$-axis (e.g. "RR, RR") labels?}
%\end{figure*}

%We also considered another approach for deriving the conclusion categories motivated by \cite{kieser2005} with the thresholds 1.11 or 1.33 but the derived conclusions are similar to the approach inspired by the methods paper of the \cite{iqwigmethodenpaper}. The main difference is that using categories motivated by \cite{kieser2005} includes two categories which are rarely occupied but using the approach motivated by the methods paper of the \cite{iqwigmethodenpaper} leads to a smaller number of empty categories.

% -----------------------------------------------------------------------------
\section{Results}
% -----------------------------------------------------------------------------

% -----------------------------------------------------------------------------
\subsection{Description of data}
% -----------------------------------------------------------------------------

Ten organisations provided 17 trials including 186 types of AEs (median 8; interquartile range [3; 9]). Twelve (71.6\% out of 17) trials were from oncology, nine (52.9\%) were actively controlled and eight (47.1\%) were placebo controlled. Median follow-up was 927 days in Arm E (interquartile range [449; 1380]), 896 days in Arm C (interquartile range [308; 1263]), and 856 days (interquartile range [308; 1234]) in both arms combined. The trials included between 200 and 7171 patients (median: 443; interquartile range [411; 1134]). In Arm E, estimated values from the Aalen-Johansen estimator ranged from 0 to 0.95 with a median of 0.09; in Arm C the range was from 0 to 0.77 with a median of 0.05. RRs based on these estimates ranged from 0.28 to 16.81 with a median of 1.71. HRs for AE hazards (restricted to AEs with $\ge$ 10 events in each arm) ranged from 0.14 to 10.83, with a median of 1.30.

\begin{figure*}[ht]
\setkeys{Gin}{width=1\textwidth}
\begin{center}
\includegraphics{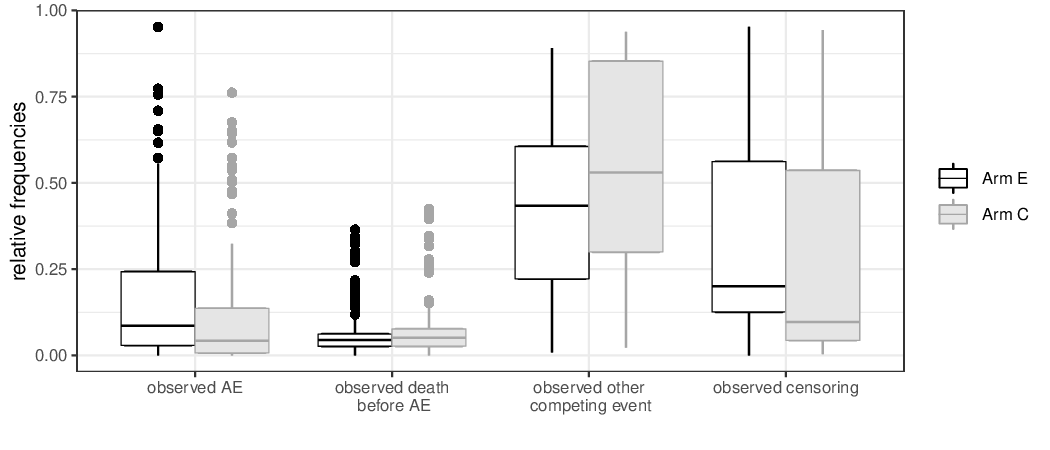}
\caption{Relative frequency of observed events, per treatment arm.}
\label{fig20:relfreq}
\end{center}
\end{figure*}

Figure~\ref{fig20:relfreq} displays for the 186 types of AEs boxplots of the treatment-arm specific observed relative frequencies, i.e., the number of {\color{black}patients with} a specific type of event divided by the total number of patients. Events considered are ``observed AE'', ``observed death before AE'', ``observed other CEs'' (i.e. excluding death), and ``observed censoring''. Within each arm, the figure illustrates a smaller amount of observed censoring compared to observed CEs. That is, AE recording often ended due to death or other CEs such as treatment discontinuation, preventing censoring of the time to AE. Comparing the arms we observe more AEs in the treatment Arm E, a comparable number of deaths, and more CEs in the control Arm C. All combined, this leads to less censoring in the control Arm C.

% -----------------------------------------------------------------------------
\subsection{Impact on RR categories and decisions based on probability-based relative AE risk estimates}
% -----------------------------------------------------------------------------

In this paragraph we summarize a key finding of this paper: namely, that categorization of evidence based on RR crucially depends on the estimator one uses to estimate the RR. Table~\ref{tab:categoriesAEprobs} shows the evidence categories for our considered estimators of the RR of those AEs where neither the estimated AE probability in Arm E nor the estimated AE probability in Arm C is 0 ($n = 155$ types of AEs for one minus Kaplan-Meier and $n = 156$ types of AEs for all other estimators).
%- n=156 (gesamt Anzahl types of AE mit werden AE prob in A=0 noch AE prob in B=0, d.h. keine 0 im RR)
%- n=155 (der 1-Kaplan-Meier ist einmal öfter AE prob =0)
%- n=94 (wegen HR: Es kam beim Cox Modell mehrere HR >20 vor. Als ich das beim Treffen in Göttingen erwähnt habe, kam der Kommentar, dass HRs mit Cox normalerweise nur berechnet werden, wenn mehr als 10 Ereignisse vorliegen.
%(Referenz die ich dazu gefunden habe:J. Concato, P. Peduzzi, T.R. Holford, A.R. Feinstein (1995)
%Importance of events per independent variable in proportional hazards analysis. I. Background, goals, and general strategy; J Clin Epidemiol, 48 , pp. 1495-1501). Deswegen sind alle Analysen zum HR auf dem Teildatensatz mit #AE in A>10 und #AE in B>10 erstellt.
Overall, we find quite a number of switches to neighboring categories. Reasons for switches are wider CIs of the Aalen-Johansen estimator as well as RR estimates / CI bounds that are close to the cutoffs between categories. As the incidence proportion on average estimates the RR well (third column in Table~\ref{tab:metaanalysis}), we see a similar number of switches to a higher ($n = 8$, below the diagonal in Table~\ref{tab:categoriesAEprobs}) and lower ($n = 9$) evidence category. Interestingly, while on average the probability transform of the incidence density accounting for CEs is approximately unbiased as well, we see double as many effect upgradings ($n = 14$) as downgradings ($n = 7$). Quite logically, for those estimators that underestimate the RR with respect to the gold-standard Aalen-Johansen estimator, namely probability transform incidence density ignoring CEs, one minus Kaplan-Meier, and Aalen-Johansen (death only) we see relevantly more switches to a lower than higher evidence category, namely $n = 41 / n = 16$, $32/8$, and $28/6$, respectively.

Switches between categories are more rare for a given estimator for earlier follow-up times, mainly because of increased variability (results not shown).

In summary, the choice of the estimator of the RR does have an impact on the conclusions.
%Furthermore, we find in general more switches than for estimation of AE probabilities in one-sample, see Table~2\kr{Verify Table number again before submission!} in Stegherr et al.\cite{onesample}.

In what follows, we will describe the different properties of the considered estimators that ultimately lead to this relevant number of diverging conclusions.

\begin{table}
\centering
 \begin{tabular}{@{}c@{}c@{}c@{}c@{}c@{}c@{}clcccc}
 \toprule
&&& &&&&& \multicolumn{4}{c}{gold-standard Aalen-Johansen}\\
&&&&&&& &(0) no effect & (a) minor & (b) considerable & (c) major \\
\midrule
&&&     \multirow{4}*{\rotatebox{90}{incidence}} & \multirow{4}*{\rotatebox{90}{proportion}}
&&&     (0) no effect    &\textbf{84} & 5           & \ndz        & \ndz        \\
&&&&&&& (a) minor        & 3          & \textbf{10} & 2           & \ndz        \\
&&&&&&& (b) considerable & 1          & 2           & \textbf{12} & 2           \\
&&&&&&& (c) major        & 1          & \ndz        & 1           & \textbf{33} \\
\midrule
         \multirow{4}*{\rotatebox{90}{probability}}&\multirow{4}*{\rotatebox{90}{transform}}&\multirow{4}*{\rotatebox{90}{incidence}}& \multirow{4}*{\rotatebox{90}{density}} &\multirow{4}*{\rotatebox{90}{ignoring }} & \multirow{4}*{\rotatebox{90}{CE}}
&&     (0) no effect    &\textbf{73} & 13          & 9           & 4           \\
&&&&&&& (a) minor        & 11         & \textbf{2}  & 5           & 3           \\
&&&&&&& (b) considerable & 2          & 2           & \textbf{1}  & 7           \\
&&&&&&& (c) major        & 1          & \ndz        & \ndz        & \textbf{21} \\
\midrule
&\multirow{5}*{\rotatebox{90}{one}}&\multirow{5}*{\rotatebox{90}{minus}}&     \multirow{5}*{\rotatebox{90}{ Kaplan-}}& \multirow{5}*{\rotatebox{90}{Meier}}
&&&     (0) no effect    &\textbf{84} & 9           & 4           & 8           \\
&&&&&&& (a) minor        & 3          & \textbf{6}  & 3           & 3           \\
&&&&&&& (b) considerable & 2          & 1           & \textbf{7}  & 5           \\
&&&&&&& (c) major        & \ndz       & 1           & 1           & \textbf{18} \\
\midrule
         \multirow{4}*{\rotatebox{90}{probability}}&\multirow{4}*{\rotatebox{90}{transform}}&\multirow{4}*{\rotatebox{90}{incidence}}& \multirow{4}*{\rotatebox{90}{density}} &\multirow{4}*{\rotatebox{90}{accounting}}&\multirow{4}*{\rotatebox{90}{for CE}}
&&     (0) no effect    &\textbf{77} & 3           & \ndz        & \ndz        \\
&&&&&&& (a) minor        & 10         & \textbf{12} & 3           & 1           \\
&&&&&&& (b) considerable & 1          & 2           & \textbf{12} & \ndz        \\
&&&&&&& (c) major        & 1          & \ndz        & \ndz        & \textbf{34} \\
\midrule
&&      \multirow{5}*{\rotatebox{90}{Aalen-}}&  \multirow{5}*{\rotatebox{90}{Johansen}}& \multirow{5}*{\rotatebox{90}{(death}} &\multirow{5}*{\rotatebox{90}{only)}}
&&     (0) no effect    &\textbf{86} & 9           & 4           & 5           \\
&&&&&&& (a) minor        & 3          & \textbf{6}  & 2           & 2           \\
&&&&&&& (b) considerable & \ndz       & 2           & \textbf{8}  & 6           \\
&&&&&&& (c) major        & \ndz       & \ndz        & 1           & \textbf{22} \\
\bottomrule
\end{tabular}
\caption{The impact of the choice of relative effect estimator for AE probabilities on qualitative conclusions. Diagonal entries are set in bold face. Deviations from the gold-standard Aalen-Johansen estimator are the off-diagonal entries. Off-diagonal zeros are omitted from the display.}
\label{tab:categoriesAEprobs}
\end{table}

% -----------------------------------------------------------------------------
\subsection{Estimators of AE probabilities compared to gold-standard}
%\kr{Tim: "For my taste the AE risks take to much space here. It takes quite
%  long to get to the actual two arm comparison." Do you consider this now to
%  be fixed, given that we have shifted the categorization paragraph to the
%  beginning?}\\{\color{red}Jan hier: Ich finde die neue Reihenfolge und die
%  Motivation \emph{In what follows, we will describe the different properties
%    of the considered estimators that ultimately lead to this relevant number
%    of diverging conclusions.} gut! Ich habe den folgenden Abschnitt noch
%  gekuerzt. Bitte sorgfaeltig schauen, ob ich keinen Bockmist gemacht habe und
%Relevantes oder Notwendiges versehentlich rausgeworfen habe.}
% -----------------------------------------------------------------------------

\begin{figure*}[ht]
\setkeys{Gin}{width=1\textwidth}
\begin{center}
\includegraphics{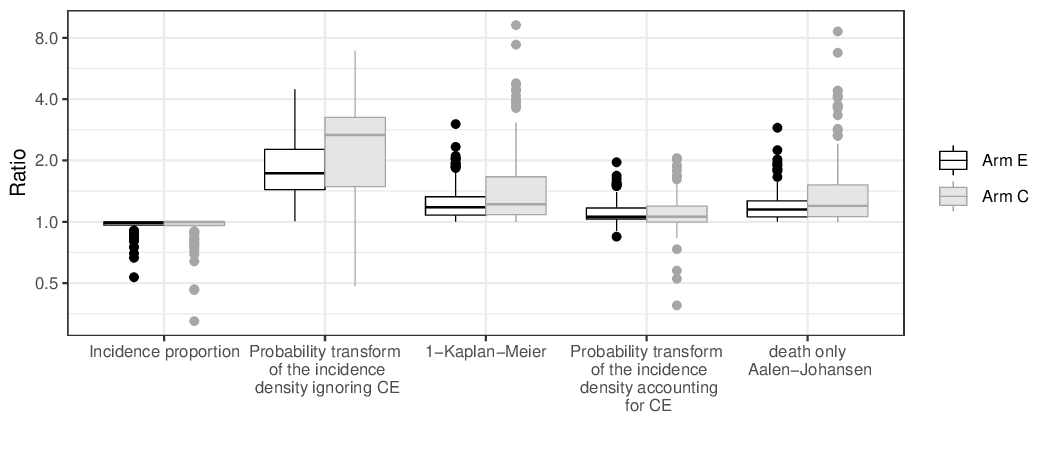}
\caption{Ratios of one-sample estimators per treatment arm. The denominator is always the gold-standard Aalen-Johansen estimator.}
%Estimated average ratio of the meta-analyses of the log ratio of the estimator of interest divided by the gold-standard Aalen-Johansen estimator from Table~\ref{tab:metaanalysis} (third column) added as gray squares.}
\label{fig30:ratios}
\end{center}
\end{figure*}

Figure~\ref{fig30:ratios} shows box plots of the ratio of the one-sample
estimators defined earlier divided by the gold-standard Aalen-Johansen
estimator, separately per treatment arm. Boxplots for Arm E
  are slightly different compared to Stegherr et al. \cite{onesample}, because we use here
  $\tau_{max}$ as evaluation time. Briefly, incidence proportion and
  gold-standard often perform comparably. Probability transforms of the incidence density perform worst when
  ignoring CE, but when accounting for CE perform much better than the other
  three procedures which are clearly biased with many examples of extreme
  overestimation. Also, the incidence proportion displays examples of biases
  (downwards), with underestimation of up to 67\%. Comparing the two arms,
  overestimation of the AE probability is more pronounced in Arm C.
  %The reason are more observed CEs in Group~B, see Figure~\ref{fig20:relfreq}, which are wrongly censored\kr{Potentially update after revision of 1-sample paper.}.
  These biases become less pronounced when looking
  at earlier evaluation times which prevent CEs and censoring after
  the respective horizon to enter calculations (results not shown).

\subsection{Meta-analysis for estimators of AE probabilities, per arm}
\label{sec:metaregAEprob1}
% -----------------------------------------------------------------------------

Meta-analyses of all estimators divided by the gold-standard Aalen-Johansen estimator are displayed in the first two columns of
Table~\ref{tab:metaanalysis}. These results confirm the visual impression
gathered from the boxplots in Figure~\ref{fig30:ratios}, but we note that
Figure~\ref{fig30:ratios} also displays biases much more pronounced than the
meta-analytical averages. In general, the amount of overestimation increased
with later evaluation times (results for earlier evaluation times not shown). Further
  investigations included uni- and multivariable meta-regressions, see the
  Stegherr et al. \cite{onesample} for Arm E results. Results for Arm C
reflect the different event pattern described above and are consistent (data
not shown).

\begin{sidewaystable*}[h]
\centering
\begin{tabular}{lcc|c}
\toprule
	                                                        & Experimental        & Control              & Ratio of RR with 95\% CI      \\
\midrule
Incidence Proportion                                        & 0.974 [0.966;0.982] & 0.978 [0.970; 0.985] & 0.997 [0.991; 1.002] \\
Probability Transform of the Incidence Density ignoring CE  & 1.817 [1.733;1.904] & 2.424 [2.249; 2.613] & 0.732 [0.703; 0.763] \\
One minus Kaplan-Meier 	                                    & 1.187 [1.161;1.214] & 1.321 [1.257; 1.389] & 0.838 [0.786; 0.894] \\
Probability Transform Incidence Density Accounting for CE   & 1.099 [1.080;1.118] & 1.124 [1.093; 1.156] & 0.977 [0.957; 0.997] \\
Aalen-Johansen (death only)                                 & 1.146 [1.125;1.168] & 1.254 [1.201; 1.308] & 0.860 [0.811; 0.911] \\
\bottomrule
\end{tabular}
\caption{Results of the meta-analyses of the log ratio of the estimator of interest divided by the gold-standard Aalen-Johansen estimator. The first two columns show the estimated average ratio and 95\% CI per treatment arm and meta-analyze the data shown in Figure~\ref{fig30:ratios}. The third column gives results of the meta-analyses of the response variable {\it log ratio of the RRs estimated with the estimator of interest and the gold-standard Aalen-Johansen estimator}, the estimated average ratio and 95\% CI. The denominator is the RR obtained using the gold-standard Aalen-Johansen estimator. This third column relates to the Panel A in Figure~\ref{fig40:comparison_rr}.}
\label{tab:metaanalysis}
\end{sidewaystable*}

\clearpage

\subsection{Estimators of relative AE risk based on probabilities compared to gold-standard}
% -----------------------------------------------------------------------------

\begin{figure*}[b]
\setkeys{Gin}{width=1\textwidth}
\begin{center}
\includegraphics{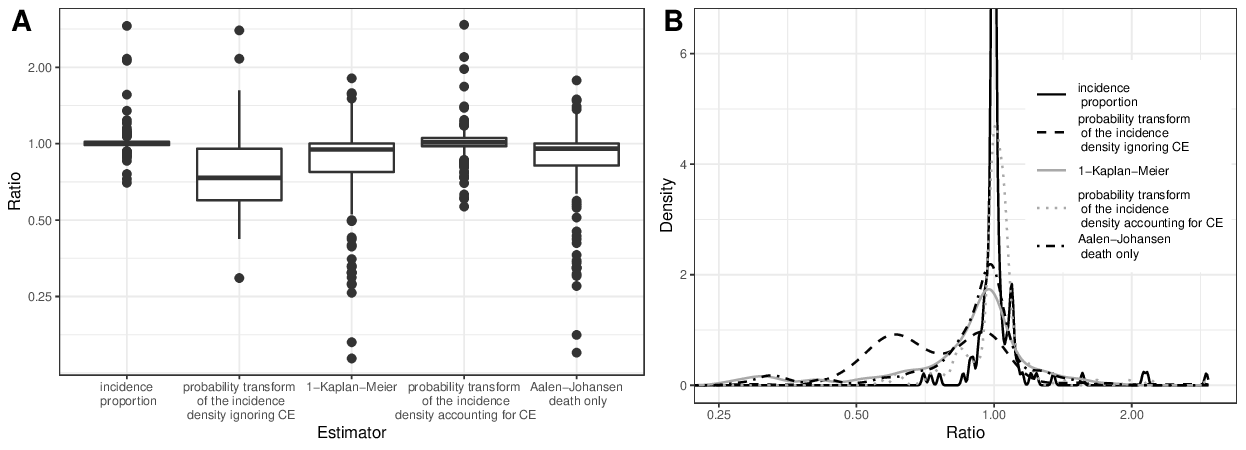}
\caption{Panel A: Ratio of RRs estimated with estimator of interest and the gold-standard Aalen-Johansen estimator. Panel B: Kernel density estimates of the RR based on AE probabilities of the estimators divided by the gold-standard Aalen-Johansen estimator.}
\label{fig40:comparison_rr}
\end{center}
\end{figure*}

Panel A in Figure~\ref{fig40:comparison_rr} displays boxplots of ratios of RRs estimated with estimator of interest and the gold-standard Aalen-Johansen estimator. Interestingly, dividing the two biased estimates of the AE probability based on the incidence proportion, which both tend to {\it underestimate} the true AE probability, leads to an estimate of the RR that on average performs comparably to the Aalen-Johansen estimator. However, note that compared to the latter, we see instances with an overestimation of the RR of just short of a factor 3. Performance is comparable for one minus Kaplan-Meier and Aalen-Johansen (death only): they both overestimate arm-specific AE probabilities but generally underestimate the RR compared to the gold-standard. This is even more pronounced for the probability transform of the incidence density ignoring CEs: it overestimates AE probabilities most, resulting in generally largest underestimation of the RR compared to the gold-standard. Finally, the probability transform of the incidence density accounting for CEs has a performance comparable to the incidence proportion for estimation of RR. This, in spite of quite different patterns for estimation of AE probabilities as displayed in Figure~\ref{fig30:ratios}. Apart from shedding light on estimation quality of individual estimators the latter is a key conclusion from comparing Figures~\ref{fig30:ratios} and \ref{fig40:comparison_rr} (Panel A): different patterns of under- or overestimation of AE probabilities can lead to similar performance for RR.
%To make that more explicit at least for averages, we have added as gray squares to Figure~\ref{fig30:ratios} the estimated average ratios of the meta-analyses of the log ratio of the estimator of interest divided by the gold-standard Aalen-Johansen estimator from the third column of Table~\ref{tab:metaanalysis}.
This implies that in general, one cannot conclude how an estimator of the relative AE risk performs based on looking how these same estimators performs on estimation of arm-wise AE probabilities.

It is interesting to see that the estimators estimating a higher AE probability than the gold-standard Aalen-Johansen estimator, namely probability transform incidence density ignoring CEs, one minus Kaplan-Meier, and Aalen-Johansen (death only), yield a smaller RR compared to the gold-standard. This is because all these estimators do not appropriately take into account CEs and overestimate the AE probability the more CEs in an arm there are. Since we have more CEs in Arm C, this eventually leads to an underestimation of the RR. The incidence proportion and the probability transform incidence density accounting for CEs correctly deal with CEs, leading on average to good estimation performance of the RR.

As discussed in Stegherr et al. \cite{onesample} one reason for the good performance of the incidence proportion might be a high amount of CEs before possible censoring. However, not only the proportion of censoring but also the timing of the censoring are relevant, as illustrated in Stegherr et al.

Factors that influence the respective behaviour of a given estimator are discussed below.

% -----------------------------------------------------------------------------
\subsection{Meta-analyses for estimators of relative AE risk based on probabilities}
\label{sec:metaregRRprob1}
% -----------------------------------------------------------------------------

The third column in Table~\ref{tab:metaanalysis} displays the results of the random-effects meta-analyses for the log ratio of the two RRs. These are in line with the results discussed above. The average ratio between the RR calculated with the incidence proportion and RR calculated with the Aalen-Johansen estimator is close to 1. The biggest underestimation is observed for the probability transform incidence density ignoring CE, with an average difference of the risks of about 27\%. The Aalen-Johansen (death only) estimator has a more pronounced underestimation compared to its counterpart accounting for all CE, but this reduction is not as pronounced as the one using the one minus Kaplan-Meier or the probability transform of the incidence density ignoring CEs. Finally, the probability transform of the incidence density accounting for CEs only slightly more underestimates on average compared to the incidence proportion. In general, these differences decrease when considering earlier evaluation times (data not shown).

%\begin{table*}[h]
%\small\sf\centering
%\begin{tabular}{lcc}
%\toprule
%Estimator                                                 & RR with 95\% CI \\
%\midrule
%Incidence Proportion                                      &	0.997 [0.991; 1.002]  \\
%Probability Transform Incidence Density                   &	0.732 [0.703; 0.763]  \\
%One minus Kaplan-Meier 	                                  & 0.838 [0.786; 0.894]  \\
%Probability Transform Incidence Density Accounting for CE &	0.977 [0.957; 0.997]  \\
%Aalen-Johansen (death only)	                              & 0.860 [0.811; 0.911] \\
%\bottomrule
%\end{tabular}
%\caption{Results of the meta-analyses of the response variable {\it log ratio of the RRs estimated with the estimator of interest and the gold-standard Aalen-Johansen estimator}. Estimated average ratio and 95\% CI. The denominator is the RR obtained using the gold-standard Aalen-Johansen estimator.}
%%Die Zielgroesse in der Meta-analyse ist z.B. log(RR IP/RR AJE). Die Ergebnisse in der Tabelle sind dann die exp-transformierten Schaetzer, stellen also z.B. RR IP/RR AJE dar.
%% Also die meta-analyse wird auf der log-Skala durchgeführt und exp-transformierte Schaetzer (estimated average ratio) dargestellt, weil die Interpretation von dieser direkter ist.
%\label{tab:metaanalysisRR}
%\end{table*}

% -----------------------------------------------------------------------------
\subsection{Uni- and multivariable meta-regressions for estimators of relative AE risk based on probabilities}
\label{sec:metaregRRprob2}
% -----------------------------------------------------------------------------
%\kr{Tim: "Diesen Abschnitt wuerde ich dahingehend kuerzen, dass ich nichts mehr ueber die average RR schreiben wuerde, da die schon im vorhergehenden Abschnitt besprochen sind. Ich wuerde mich hier auf die Effekte %der covariates konzentrieren."}

For the meta-regressions, we use as covariates the percentage of censoring (for both arms combined), the percentage of CEs (for both arms combined), the maximum follow-up time, and the size of the RR as estimated by the gold-standard Aalen-Johansen estimator. Covariates were centered, i.e., the row ``average RR'' contains the average RR of the estimator of interest and the Aalen-Johansen estimator if the covariate takes its mean. These means were 28.6\% censoring (= percentage of censored observations until $\tau_{max}$), 53.8\% CEs, 2.38 years (781 days) evaluation time, and a RR of 2.55. Table~\ref{tab:metaregressionuniresRR} provides results from the uni- and Table~\ref{tab:metaregressionmultiresRR} from the multivariable analysis.

%Overall, and in line with the results in the third column of Table~\ref{tab:metaanalysis}, all estimators had an estimated average RR below one, i.e. a lower average RR compared to the gold-standard Aalen-Johansen %estimator. There are hardly any differences between the RR calculated with the incidence proportion and the RR calculated with the Aalen-Johansen estimator in meta-regression with a single independent variable.
We illustrate the interpretation of the parameters of the meta regression models by the following example calculation: The average ratio of the RR calculated with the probability transform of the incidence density ignoring CEs and the RR calculated with the Aalen-Johansen estimator at $\tau_{max}$ under 28.6\% censoring is 0.729. If a trial has 38.6\% censoring, i.e. an increased censoring proportion of 10 percentage points, the average ratio is estimated as $0.729 \cdot 1.066 = 0.777$.

%Results for the univariable analysis of the incidence proportion and the probability transform of the incidence density accounting for CEs are comparable in that their average RR is virtually the same as the one for %the gold-standard Aalen-Johansen estimator, and the impact of any of the covariates is very small, cf. also Figure~\ref{fig40:comparison_rr}.

The biggest average underestimation of the RR compared to the gold-standard is seen for the probability transform of the incidence density ignoring CEs (27 percentage points) and about the same for one minus Kaplan-Meier and Aalen-Johansen (death only, both about 16 percentage points). Increasing the censoring proportion compared to their respective mean leads to less underestimation with respect to the gold-standard while the opposite is true for increasing the proportion of CEs. Increasing the evaluation time or the size of the gold-standard RR compared to their respective mean has no relevant additional effect. For the probability transform incidence density accounting for CEs results are on average comparable to the gold-standard.

These results are confirmed in the multivariable analysis in Table~\ref{tab:metaregressionmultiresRR}.

In summary, these meta-regressions show that (1) the key difference between estimators lies in the value of the average RR and (2) the impact of covariates is overall limited, compared to the average RR the estimated coefficients are close to 1. This emphasizes that the choice of the estimator is key, and that this holds true over a wide range of possible data configurations quantified through the considered covariates.

\begin{sidewaystable}
\footnotesize
\centering
    \begin{tabular}{llccccc}
 \toprule
&& & probability transform& & probability transform & \\
&&incidence &incidence density&one minus& incidence density & Aalen-Johansen\\
&&proportion & ignoring CE & Kaplan-Meier & accounting for CE & (death only) \\
\midrule
\% censoring &average RR&  0.995 [0.989; 1.001] & 0.729 [0.706; 0.752] & 0.828 [0.779; 0.881] & 0.977 [0.957; 0.997] & 0.851 [0.805; 0.900] \\
&10\% increase & 1.001 [0.999; 1.004] & 1.066 [1.053; 1.080] & 1.047 [1.023; 1.071] & 0.998 [0.990; 1.006] & 1.040 [1.018; 1.062] \\
\midrule
\% CEs &average RR& 0.996 [0.990; 1.002] & 0.719 [0.697; 0.741] & 0.817 [0.768; 0.870] & 0.981 [0.961; 1.001] & 0.842 [0.795; 0.891] \\
&10\% increase  & 0.999 [0.996; 1.002] & 0.934 [0.922; 0.946] & 0.950 [0.927; 0.973] & 1.012 [1.003; 1.020] & 0.959 [0.938; 0.980] \\
\midrule
size of RR &average RR& 0.998 [0.991; 1.006] & 0.730 [0.702; 0.760] & 0.837 [0.786; 0.891] & 0.977 [0.958; 0.998] & 0.859 [0.812; 0.910] \\
&increase of 0.1& 1.000 [1.000; 1.001] & 0.997 [0.996; 0.999] & 0.997 [0.994; 0.999] & 1.001 [1.000; 1.001] & 0.997 [0.995; 0.999] \\
\midrule
evaluation time&average RR & 0.995 [0.989; 1.001] & 0.728 [0.699; 0.758]& 0.838 [0.785; 0.894] &0.978 [0.958; 0.999] &0.860 [0.811; 0.911]\\
&one additional year & 1.004 [1.000; 1.001] & 1.033 [1.002; 1.006] & 1.006 [0.959; 1.005] & 0.992 [0.978; 1.001] & 1.000 [0.958; 1.004] \\
\bottomrule
\end{tabular}
\caption{Average RR and multiplicative change by 10\% increase in amount of censoring, 10\% increase in CEs, one additional year of observation or a 0.1 greater RR from univariable meta-regressions. The RR is estimated by the gold-standard Aalen-Johansen estimator.}
\label{tab:metaregressionuniresRR}
\end{sidewaystable}

\clearpage

\begin{sidewaystable}
\footnotesize
\centering
        \begin{tabular}{lccccc}
 \toprule
& & probability transform& & probability transform & \\
&incidence &incidence density&one minus  & incidence density & Aalen-Johansen\\
&proportion & ignoring CE &Kaplan-Meier& accounting for CE & (death only) \\
\midrule
average RR &  0.997 [0.990; 1.005] & 0.727 [0.706; 0.750] & 0.829 [0.780; 0.880] & 0.979 [0.959; 0.999] & 0.853 [0.808; 0.901] \\
\% censoring 10\% increase & 0.999 [0.996; 1.003] & 1.068 [1.055; 1.082] & 1.053 [1.028; 1.078] & 0.999 [0.990; 1.007] & 1.045 [1.023; 1.068] \\
size of RR increase of 0.1 & 1.000 [1.000; 1.001] & 0.997 [0.996; 0.998] & 0.996 [0.994; 0.999] & 1.001 [1.000; 1.001] & 0.997 [0.995; 0.999] \\
evaluation time one additional year  & 1.005 [1.000; 1.011] & 0.987 [0.964; 1.011] & 0.973 [0.929; 1.020] & 0.994 [0.979; 1.010] & 0.973 [0.932; 1.015] \\
\bottomrule
\end{tabular}
\caption{Results of the multivariable meta-regressions of the response variable {\it log ratio of the RRs, estimated with the estimator of interest and the gold-standard Aalen-Johansen estimator} with centered covariates. The size of the RR is estimated by the gold-standard Aalen-Johansen estimator.}
\label{tab:metaregressionmultiresRR}
\end{sidewaystable}

\clearpage
% -----------------------------------------------------------------------------
\subsection{Variability for estimators of relative AE risk based on probabilities}
% -----------------------------------------------------------------------------

So far, we have primarily discussed how the different estimators perform compared to the gold-standard Aalen-Johansen estimator {\it on average}. In a first step, variability of the RRs of every estimator with respect to the gold-standard Aalen-Johansen estimator can be assessed from the switches between RR categories in Table~\ref{tab:categoriesAEprobs} and the boxplots in Panel A in Figure~\ref{fig40:comparison_rr}. Here, we provide a more detailed account of the variability of the RRs using kernel density estimates of their distribution. As in the one-sample scenario, on average the incidence proportion appears to provide a good estimator of relative AE risk based on probabilities. Considering the plot of the kernel density estimates of the RRs of the AE probability in Panel B of Figure~\ref{fig40:comparison_rr}, the RR based on the incidence proportion and the gold-standard is most often close to one. However, there is also a peak of the estimated kernel density at larger ratios, indicating that the estimators are not always comparable. For the RR based on the ratio of the probability transform of the incidence density accounting for CEs and the gold-standard we still have clustering around one, but not as pronounced as for the incidence proportion. The ratios of the one minus Kaplan-Meier or Aalen-Johansen (death only) estimator have less values close to one. For these two estimators more values are smaller than one than larger. The estimated kernel density of the probability transform of the incidence density ignoring CE has no peak at one but is bimodal with both modes below one.

Patterns of {\color{black}relative }frequencies of the AE itself, censoring, and CEs, that may lead to extreme discrepancies between a given estimator and the gold-standard Aalen-Johansen estimator for estimation of AE probabilities are discussed in detail in Stegherr et al. \cite{onesample} Of course, such extreme configurations in one or both arms may lead to extreme RR estimates also for the two arm comparison.

%\begin{figure*}[t]
%\setkeys{Gin}{width=1\textwidth}
%\begin{center}
%\includegraphics{plots/fig45_kernel_RR.png}
%\caption{Kernel density estimates of the RR based on AE probabilities of the estimators divided by the gold-standard Aalen-Johansen estimator.}
%\label{fig45:kernel_rr}
%\end{center}
%\end{figure*}

% -----------------------------------------------------------------------------
\subsection{Estimators of relative AE risk based on hazards compared to gold-standard}
% -----------------------------------------------------------------------------

\begin{figure*}[t]
\setkeys{Gin}{width=0.8\textwidth}
\begin{center}
\includegraphics{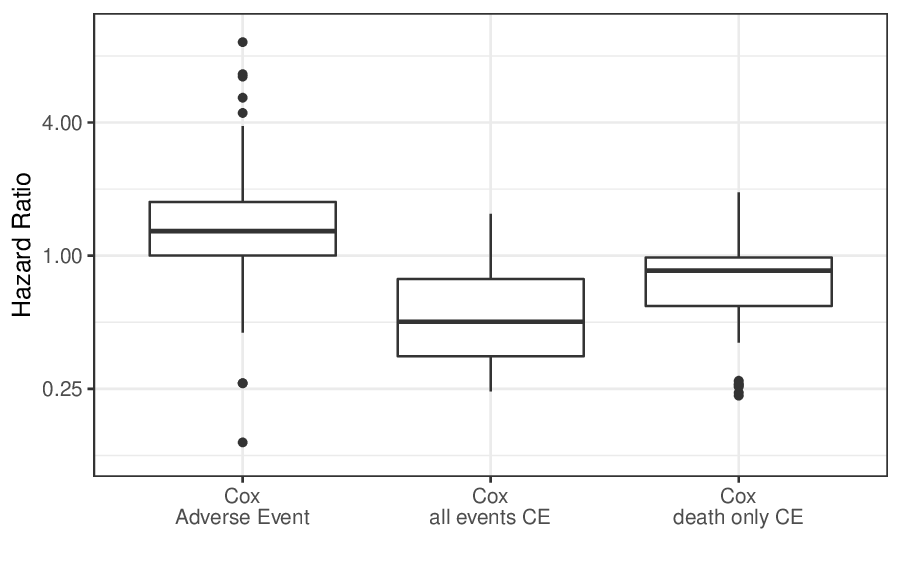}
\caption{Cox regression HRs (on log-scale) for the three event types AE, all CEs, and CE of death.}
\label{fig47:hr}
\end{center}
\end{figure*}

Figure~\ref{fig47:hr} displays boxplots of the HRs calculated from Cox regression. The three boxplots display HRs for hazards of AE, all CEs, and a CE of death. Assessing the effect for the endpoint of interest, here time to AE, as well as of any CE, here time to all CEs or time to a CE of death, is generally recommended for any (hazard-based) analysis of competing events \cite{latouche_13}. We find that the hazard of AE is generally larger for Arm E compared to Arm C, meaning that the instantaneous risk of AE is typically higher, not unexpectedly. For the hazards of CEs, for both types, what we find is that the hazard in Arm E is generally lower than in Arm C, i.e. there is an effect of the experimental treatment on the CE. If we simply censored at CEs we would thus introduce arm-dependent censoring, a feature that may lead to biased effect estimates \cite{schemper_96, clark_02}. We will use this to explain observations we make in Table~\ref{tab:RR_HR} below.

%\kr{Dies haengt erneut vom Zusammenspiel der Ueberlebensverteilungen fuer AE und CEs ab, es ist nicht so, dass dies aus der Tatsache, dass AE HR $>$ 1 FOLGT. Sollen wir das diskutieren (stimmt es ueberhaupt)?}

\begin{figure*}[!h]
\setkeys{Gin}{width=\textwidth}
\begin{center}
\includegraphics{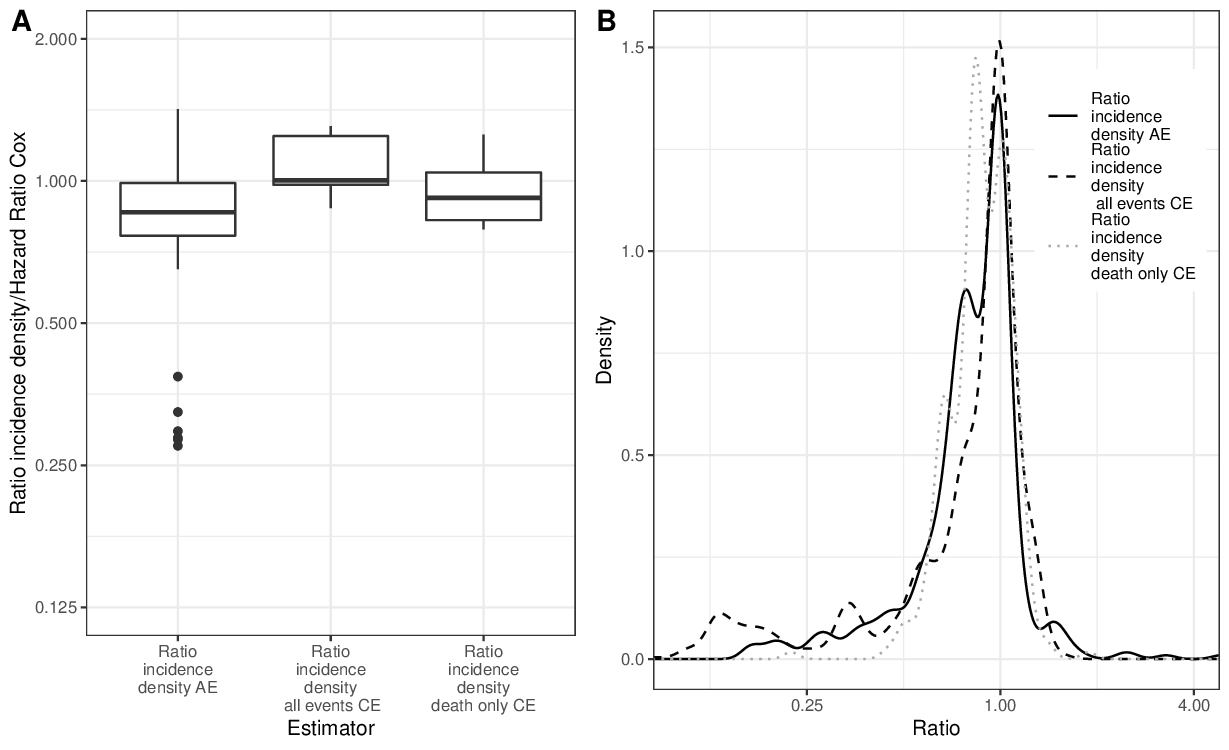}
\caption{Panel A: Ratio of RRs estimated with estimator of interest and the gold-standard HR based on Cox regression estimator, all evaluated at $\tau_E$ and $\tau_C$. Panel B: Plots of the kernel density estimates of the HR of the estimators divided by the gold-standard Cox-regression estimator.}
\label{fig50:comparison_hr}
\end{center}
\end{figure*}

Boxplots of the ratios of incidence density estimates in each arm, evaluated at $\tau_E$ and $\tau_C$, respectively, and the gold-standard HR calculated from Cox regression, are provided in Panel A in Figure~\ref{fig50:comparison_hr}, for again the same endpoints as in Figure~\ref{fig47:hr}. The ratio of the incidence densities of the AE in the two arms underestimates with respect to the Cox regression HR while for the other two endpoints on the median they turn out to be approximately unbiased compared to the Cox HR, with a tendency to overestimation when accounting for all CEs and underestimation when only accounting for a CE of death.

% \kr{Ich hab mal versucht, die relevanten Aspekte fuer eine Interpretation zusammenzutragen, auch, um die Meta-Analyse Resultate weiter unten besser zu verstehen:}
To appreciate the differences between the two estimators of the RR based on hazards, i.e. the incidence density ratio and the gold-standard Cox regression HR, recall the properties of the two methods: Both properly account for censoring and they properly estimate event-specific hazards, or rather the relative effect based on these. The only difference between the two methods is what they assume about the shape of the underlying hazard: the incidence density assumes them to be {\it constant} up to the considered follow-up time, which also implies that they are {\it proportional}. The gold-standard Cox regression HR only assumes them to be {\it proportional}, but not constant. But in addition, we also have different data patterns for the three event causes for which we show boxplots of relative hazard-based risks in Panel A in  Figure~\ref{fig50:comparison_hr}: looking at Figure~\ref{fig20:relfreq} we find a low proportion of events for AE and an even lower proportion of CEs of death, but a higher proportion of events for all CEs. Furthermore, within a given patient, death happens later than AE. So the results we observe in Panel A in Figure~\ref{fig50:comparison_hr} are a result of different tradeoffs between all these aspects.

If we restrict follow-up to earlier timepoints, then variability increases and on average, results persist for the events of AE and CE of death, but for all CEs the ratio of incidence densities underestimates with earlier timepoints (data not shown). At earlier timepoints the number of events for all CEs also decreases, so that the tradeoff with the constant hazard assumption starts to resemble that of an event of AE.

% -----------------------------------------------------------------------------
\subsection{Meta-analyses for estimators of relative AE risk based on hazards}
% -----------------------------------------------------------------------------

Table~\ref{tab:metaanalysisHR} confirms these findings in univariable meta-analyses and provides an average quantification of the amount of underestimation of all three estimators relative to the Cox regression HR.

\begin{table*}[!h]
\centering
\begin{tabular}{lcc}
\toprule
Estimator                                                 & ratio with 95\% CI \\
\midrule
Ratio incidence density AE                                &	0.803 [0.741; 0.871]  \\
Ratio incidence density all events CE                     &	0.908 [0.851; 0.969]  \\
Ratio incidence density (death only) CE                   & 0.958 [0.934; 0.982]  \\
\bottomrule
\end{tabular}
\caption{Results of the meta-analyses of the response variable {\it log ratio of the ratio of the estimator of interest and the HR estimated by the Cox model}. Estimated average ratio and 95\% CI. The denominator is the HR obtained using the Cox model.}
% Nein, in der meta-regression ist die Zielgröße z.B. log( (incidence density of AE in A/incidence density of AE in B)/ HR of Cox). Die Ergebnisse in der Tabelle sind dann wieder exp- transformiert um Aussagen direkt % ueber Ratio incidence density / HR of Cox treffen zu koennen.
\label{tab:metaanalysisHR}
\end{table*}

% -----------------------------------------------------------------------------
\subsection{Uni- and multivariable meta-regressions for estimators of relative AE risk based on hazards}
% -----------------------------------------------------------------------------

For the meta-regressions reported in Tables~\ref{tab:metauniHR} and \ref{tab:metamultiHR}, in line with what has been done for RR in Tables~\ref{tab:metaregressionuniresRR} and \ref{tab:metaregressionmultiresRR}, we again use as covariates the percentage of censoring, the percentage of CEs, the maximum follow-up time, and the size of the HR as estimated by the gold-standard Cox regression as covariates. Means at which covariates were centered were 29.6\% censoring (= mean percentage of censored observations until $\tau_E$ and $\tau_C$), 56.7\% CEs, 2.72 years (994 days) maximum follow-up time, and a HR of 1.74. Note that these are slightly different from the ones reported above because we are using a different follow-up time here. In the univariable analyses in Table~\ref{tab:metauniHR}, we find an estimated average RR below one for time to AE and time to a CE of death, i.e. a lower average RR compared to the gold-standard Cox regression estimator, for all covariates. For time to all CEs the ratio of incidence densities overestimates on average. Most pronounced covariate effects are found for time to AE, for example the average ratio of the HR based on the incidence density and the gold-standard Cox regression HR under 29.6\% censoring is 0.825. If a trial has 39.6\% censoring, i.e. an increased censoring proportion of 10 percentage points, the estimated average ratio is estimated as $0.825 \cdot 1.052 = 0.868$.

These results are confirmed in the multivariable analysis in Table~\ref{tab:metamultiHR}.

Similar to the estimation of RRs for AE probabilities we find that the effect of covariates compared to the estimated average HR is rather limited for all three endpoints. This means again that the choice of estimator, which either allows for an unspecified, freely varying baseline hazard (Cox) or assumes it to be constant (incidence densities), appears to be more relevant than the configuration of the data as captured by the covariates.

\begin{table*}
\centering
\begin{tabular}{llccc}
\toprule
& & incidence density&  incidence density& incidence density\\
&&of AE  & of all events CE& of death only CE\\
\midrule
\% censoring &average HR & 0.825 [0.784; 0.868] & 1.063 [1.045; 1.081] & 0.962 [0.938; 0.986] \\
&10\% increase  & 1.052 [1.031; 1.073] & 0.990 [0.983; 0.996] & 1.026 [1.015; 1.036] \\
\midrule
\% CEs &average HR& 0.786 [0.741; 0.834] & 1.061 [1.044; 1.079] & 0.942 [0.913; 0.971] \\
&10\% increase  & 0.948 [0.925; 0.971] & 1.011 [1.004; 1.017] & 1.041 [1.013; 1.071] \\
\midrule
size of HR &average HR& 0.836 [0.790; 0.884] & 1.063 [1.048; 1.078] & 0.959 [0.932; 0.987] \\
&increase of 0.1& 1.004 [1.000; 1.008] & 0.972 [0.966; 0.977] & 0.996 [0.987; 1.005] \\
\midrule
evaluation time&average HR & 0.854 [0.809; 0.902] & 1.053 [1.040; 1.067] & 0.976 [0.944; 1.010] \\
&one additional year & 0.937 [0.905; 0.970] & 0.945 [0.936; 0.955] & 0.977 [0.954; 1.000] \\
\bottomrule
\end{tabular}
\caption{Univariable meta-regression for the response variable {\it log ratio of the HRs, estimated with the estimator of interest and the gold-standard Cox regression}. The size of the HR is estimated by the Cox model.}
\label{tab:metauniHR}
\end{table*}

\begin{table*}
\centering
\begin{tabular}{lccc}
\toprule
& incidence density&  incidence density& incidence density\\
&of AE  & of all events CE& of death only CE\\
\midrule
average HR & 0.846 [0.809; 0.884] & 1.054 [1.042; 1.066] & 1.003 [0.979; 1.029] \\
\% censoring 10\% increase &  1.058 [1.040; 1.076] & 1.004 [0.999; 1.009] & 1.038 [1.029; 1.048] \\
size of HR increase of 0.1 & 1.005 [1.002; 1.008] & 0.980 [0.974; 0.986] & 0.993 [0.986; 0.999] \\
evaluation time one additional year & 0.933 [0.907; 0.960] & 0.957 [0.948; 0.966] & 0.949 [0.931; 0.968] \\
\bottomrule
\end{tabular}
\caption{Multivariable meta-regression. The size of the HR is estimated by the Cox model. Note that for the incidence density of death only CE the percentage of CEs correspond to the percentage of deaths and for the other two estimators its the percentage of all events CEs.}
\label{tab:metamultiHR}
\end{table*}

\clearpage

% -----------------------------------------------------------------------------
\subsection{Variability in estimation}
% -----------------------------------------------------------------------------

As for the estimation of RRs based on probabilities, we provide a more
detailed account of the variability of the HRs using kernel density estimates
of their distribution. Considering the plot of the kernel density estimates of
the ratios of the HRs in Panel B in Figure~\ref{fig50:comparison_hr}, the ratio of the HRs for time to AE
has its highest peak just below one, but also further peaks that are even
smaller than one, indicating that the estimators are not always
comparable. The estimated density for the ratio of HRs for time to a CE of death is multimodal,
with the highest peak further below one than for time to AE. % The higher number
% of events for time to all CEs leads to %% Jan: Ich weiss nicht, ob das so
% sein muss, higher number => closer to gold standard
A higher proportion of ratios of HRs for time to all CEs is close to one, but also this estimated density is multimodal. All densities are left-skewed, indicating that there is a relevant portion of AEs for which the ratio of incidence densities underestimates compared to the gold-standard Cox regression HR for time to all CEs.

%\begin{figure*}[t]
%\setkeys{Gin}{width=1\textwidth}
%\begin{center}
%\includegraphics{plots/fig60_kernel_HR.png}
%\caption{Plots of the kernel density estimates of the HR of the estimators divided by the gold-standard Cox-regression estimator.}
%\label{fig60:kernel_hr}
%\end{center}
%\end{figure*}

% -----------------------------------------------------------------------------
\subsection{Impact on RR categories and decisions based on hazard-based relative AE risk estimates}
% -----------------------------------------------------------------------------

%%\kr{Das habe ich jetzt mal hier hinten gelassen und nicht nach vorne geschoben. Einverstanden?}

Table~\ref{tab:RR_ID_HR} provides a comparison of the conclusions drawn from the estimators of a RR for time to AE. The majority of AEs either lead to ``no effect'' or an effect of ``major'', and these are quite consistently detected by the two methods. However, we also observe for $19 / 94 = 20.2\%$ of AEs a diverging conclusion, following from the combination of bias and variability in estimation described above.

\begin{table*}[!h]
\centering
\begin{tabular}{llcccc}
\toprule
                                                   &                  & \multicolumn{4}{@{}c@{}}{{HR Cox for AE}}              \\
                                                   &                  &(0) no effect & (a) minor  & (b) considerable & (c) major   \\
\midrule
\multirow{4}{*}{\shortstack{RR incidence\\ density}} & (0) no effect    & \textbf{49}  & 1          & \ndz             & 1           \\
                                                   & (a) minor        & 3            & \textbf{5} & 5                & \ndz        \\
                                                   & (b) considerable & 2            & \ndz       & \textbf{3}       & 1           \\
                                                   & (c) major        & 3            & \ndz       & 3                & \textbf{18} \\
\bottomrule
\end{tabular}
\caption{Conclusions of the RR calculated with the ratio of incidence densities and the HR calculated with the Cox model, both at $\tau_E,\tau_C$. The table shows the analysis of those $n=94$ AEs that were observed with a{\color{black}n absolute} frequency of $\ge 10$ in each arm. Diagonal entries are set in bold face. Divergent decisions between the two methods are the off-diagonal entries. Off-diagonal zeros are omitted from the display.}
\label{tab:RR_ID_HR}
\end{table*}

% -----------------------------------------------------------------------------
\subsection{Comparison of qualitative decisions on relative effect based on the two gold-standards}
% -----------------------------------------------------------------------------

%%\kr{Das habe ich jetzt mal hier hinten gelassen und nicht nach vorne geschoben. Einverstanden?}

We have considered two effect measures to quantify the RR of an AE in two
arms: the RR based on AE probabilities evaluated at $\tau_{max}$ and the HR with maximum available follow-up in both arms, where Cox' method of estimating the HR implicitly leads to an evaluation at $\tau_{max}$, too. Our analyses reveal that all the considered estimators
are overall inferior to the two gold-standards we considered, either the RR
based on the arm-wise Aalen-Johansen estimator or the HR based on Cox
regression. One question that remains is whether the qualitative conclusions
drawn based on the two gold-standards are relevantly different when relying on
the criteria put forward by the IQWiG (Table 14 in their general methods document \cite{iqwigmethodenpaper}). Table~\ref{tab:RR_HR} has the results. We observed quite some different
classifications based on the two estimates of the RR. However, this
is not a surprise, as the {\it estimand} the two methods look at is not the
same (see \cite{varadhan_10} for an exposition of this issue): Cox HR quantifies a relative effect based on an endpoint of AE
\emph{hazard}, whereas RR based on gold-standard Aalen-Johansen
is based on a comparison of \emph{probabilities} at a evaluation
time. The latter integrates the hazard for the endpoint of interest and the hazard for CE into one cumulative effect
measure, whereas a Cox regression only considers one hazard at a time, and
this is likely the primary reason for the divergent decisions in
Table~\ref{tab:RR_HR}. Empirically, if the boxplot for the HR for the CE in Figure~\ref{fig47:hr} would center around one, then (ignoring the fact that the categorization also takes into account uncertainty) in theory the decision based on RR and HR would approximately coincide, i.e. we would have no non-diagonal entries in Table~\ref{tab:RR_HR}. But whenever there is an effect on the CE, then it is expected that decisions diverge.

Summarizing all these analyses concerning effect quantification for AEs using
Cox regression based HRs, we conclude that the ratio of incidence densities
cannot be considered a uniformly good approximation of the HR based on Cox
regression. We also find that categorization of the relevance of differences between treatment arms may differ depending on whether it is based on one
  event-specific hazard alone (Cox for AE) or on a proper probability
  estimator (Aalen-Johansen, integrating the two event-specific hazards).

% \begin{table*}[h]
% \centering
% \begin{tabular}{llcccc}
% \toprule
%                                             &                  & \multicolumn{4}{@{}c@{}}{{RR Aalen-Johansen}}              \\
%                                             &                  &(0) no effect & (a) minor  & (b) considerable & (c) major   \\
% \midrule
% \multirow{4}{*}{\shortstack{HR Cox for AE}} & (0) no effect    & \textbf{42}  & 3          & 3                & 1           \\
%                                             & (a) minor        & 9            & \textbf{2} & 1                & \ndz        \\
%                                             & (b) considerable & 4            & 1          & \textbf{3}       & 2           \\
%                                             & (c) major        & 2            & \ndz       & 4                & \textbf{17} \\
% \bottomrule
% \end{tabular}
% \caption{Conclusions of the RR calculated with the gold-standard Aalen-Johansen estimator at $\tau_{max}$ compared to the conclusions of the HR calculated with the Cox model at $\tau_E,\tau_C$. The table shows the analysis of those $n=94$ AEs that were observed with a frequency of $>10$ in both arms.}
% \label{tab:RR_HR}
% \end{table*}

\begin{table}[h]
\centering
\begin{tabular}{llcccc}
\toprule
                                                              &                  & \multicolumn{4}{@{}c@{}}{{HR Cox for AE}}              \\
                                                              &                  &(0) no effect & (a) minor  & (b) considerable & (c) major   \\
\midrule
\multirow{4}{*}{\shortstack{RR gold-standard\\ Aalen-Johansen}} & (0) no effect    & \textbf{42}  & 3          & 3                & 1           \\
                                                              & (a) minor        & 9            & \textbf{2} & 1                & \ndz        \\
                                                              & (b) considerable & 4            & 1          & \textbf{3}       & 2           \\
                                                              & (c) major        & 2            & \ndz       & 4                & \textbf{17} \\
\bottomrule
\end{tabular}
\caption{Conclusions of the RR calculated with the gold-standard Aalen-Johansen estimator at $\tau_{max}$ compared to the conclusions of the HR calculated with the Cox model at $\tau_E,\tau_C$. The table shows the analysis of those $n=94$ AEs that were observed with a{\color{black}n absolute} frequency of $\ge 10$ in each arm. Off-diagonal zeros are omitted from the display.}
\label{tab:RR_HR}
\end{table}

% -----------------------------------------------------------------------------
\subsection{Role of censoring}
% -----------------------------------------------------------------------------

\begin{figure*}[h]
\setkeys{Gin}{width=0.7\textwidth}
\begin{center}
\includegraphics{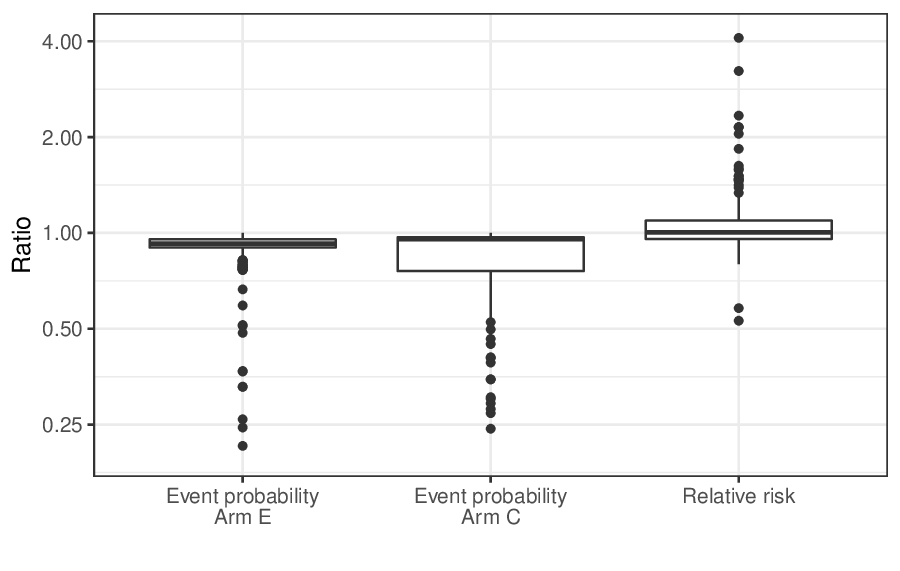}
\caption{Ratios of probability estimates (left and middle) and RR (right) based on incidence proportion of the composite endpoint combining AE and CE divided by the corresponding estimate of composite one minus Kaplan-Meier estimator.}
\label{fig46:composite}
\end{center}
\end{figure*}

In this section, we aim to explore how the incidence proportion and one minus Kaplan-Meier compare in the absence of CEs. To this end, we consider a composite endpoint where AEs and CEs are combined into one single event. The gold-standard in this setting is the one minus Kaplan-Meier estimator which is compared to the incidence proportion in Figure~\ref{fig46:composite}. Since for this endpoint we do not have CEs, the conclusions on relative effects from the probability- and hazard-based analyses are aligned in the sense of the direction of the effect.

As visible in the left and middle boxplot in Figure~\ref{fig46:composite}, in the composite endpoint analysis underestimation by the incidence proportion is more pronounced than in the analysis of the AE probability presented above. As discussed by Stegherr et al. \cite{onesample} one reason for this observation is that even in the presence of censoring, for the one minus Kaplan-Meier estimator the type of the last event is
most important. If the last event is an AE or CE the one minus Kaplan-Meier estimator is equal to one, even though censoring has been observed at earlier follow-up times. The incidence proportion is only equal to one if no censoring is observed. For the RR this leads to an overestimation compared to the gold-standard one minus Kaplan-Meier estimator.

%{\color{green} Regina: you mentioned you have a simple explanation here? If this goes beyond what we already have here, can you please add? - $RR\_IP/RR\_KM=(IP\_E/IP\_C)/(KM\_E/KM\_C)=(IP\_E/KM\_E)/(IP\_C/KM\_C)$ Unterschaetzung in Gruppe B groesser, daher Nenner kleiner als Zaehler des letzten Bruchs und deswegen ratio der RRs groesser 1. Für mich ist es daher ausreichend zu schreiben, dass es eine Konsequenz der Unterschaetzung der IP ist.}

% -----------------------------------------------------------------------------
\subsection{Role of follow-up time}
% -----------------------------------------------------------------------------

We focused on the results when evaluating estimators using the maximum follow-up time as evaluation time. When looking at earlier evaluation times where the estimators were evaluated at earlier time points
defined by quantiles of the observed times (results not shown in detail), the resulting bias was, in general, less pronounced, due to a reduced relative frequency of CEs and of censoring (see Figure~\ref{fig20:relfreq}). We regarded the situation of including all data up to the maximum follow-up time as the most relevant as this is the usual practice.

%% -----------------------------------------------------------------------------
%\subsection{Stratified results for the RR based on hazards}
%% -----------------------------------------------------------------------------
%
%\kr{We have also analyzed split by HR $< / \ge$ 1? Should we report that as well?}
%
%
%
%% -----------------------------------------------------------------------------
%\subsection{Role of Censoring for relative AE risk based on hazards}%% -----------------------------------------------------------------------------
%
%\kr{Role of Censoring ist doch IP vs KM im composite Fall. Was ist das Hazard Analogon dazu? HR vom Composite Endpoint wurde nicht berechnet...}

% -----------------------------------------------------------------------------
\section{Discussion}
% -----------------------------------------------------------------------------

Survival analyses accounting for CEs is methodologically well
established, but practical use la{\color{black}g}s behind
\cite{schumacher2016competing, austin_17a, phillips_20}. Failure to account for censoring (e.g.,
incidence proportion) or CEs (e.g., one minus Kaplan-Meier) will
generally lead to biased quantification of absolute AE risk, and the possible amount of bias has been investigated in Stegherr et al. \cite{onesample}. There, we confirmed that one minus
Kaplan-Meier should not be used to estimate the cumulative AE probability, as
it is bound to overestimate as a consequence of ignoring competing
events. Here, we found that this arm-wise overestimation often leads
to an underestimation of the RR when comparing two arms. The same pattern is
observed for the probability transform incidence density ignoring CE and
Aalen-Johansen (death only), i.e. the other two estimators that do not
correctly account for CEs.

For estimation of AE probabilities, the incidence proportion performed
surprisingly well when compared to the gold-standard Aalen-Johansen
estimator. As discussed in Stegherr et al. \cite{onesample}, the reason was a
high amount of CEs before possible censoring, potentially related
to the majority of the seventeen trials analyzed coming from oncology. These
good arm-wise estimates translate in on average unbiased estimation of RR as
well. However, as discussed by Stegherr et al., use of the incidence
proportion \emph{implicitly} assumes events to be competing as defined in the
methods section. Furthermore, although on average the incidence proportion performs well, we have still 17/156 = 10.9\% AE types that were categorized differently in Table~\ref{tab:categoriesAEprobs}, with one being turned from a ``major'' (incidence proportion) to a ``no effect'' with the Aalen-Johansen estimator.
Finally, we found that incidence densities, typically criticized because of the restrictive constant hazard assumption, led to the worst performance when their probability transform ignored CEs.
% Finally, we also confirm for RR the concerns regarding the
% constant hazard assumption underlying incidence densities\kr{Jan: "Ich lese
%   das weniger als Bestaetigung der Sorge bzgl. konst Hazards. (Die wird mE
%   mehr durch Fig~~\ref{fig50:comparison_hr} bestaetigt.)" Koenntest du das
%   entsprechend umformulieren, oder gleich loeschen?}: the probability
% transform incidence density ignoring CE performed worst.
However, accounting for CEs in an analysis that parametrically
mimicked the non-parametric Aalen-Johansen performed better than both one
minus Kaplan-Meier and Aalen-Johansen (death only). This confirms the
conclusion from the one-sample case that ignoring CEs appeared to
be worse than assuming constant hazards in our empirical study. In general, we
caution against making conclusions about the amount and direction of bias for
estimation of the RR based on the behavior of the one-sample
estimators. Overall, in terms of relevance of effects, the choice of the
estimator is key and more important than the features of the underlying data
such as percentage of censoring, CEs, amount of follow-up, or the
value of the gold-standard RR.

Kernel estimates of densities of ratios of estimated RRs by the different estimators divided by the RR estimated by the gold-standard Aalen-Johansen estimator revealed that all estimators except probability incidence density ignoring CEs have a peak just below one. However, all estimators either had further peaks away from one or the estimated density was unimodal but with high variance. This indicates that the considered estimators are not always comparable to the gold-standard Aalen-Johansen estimator. In general, it is not obvious what feature of the data generating mechanism actually leads to observed data for which, e.g., the incidence proportion performs much worse than the Aalen-Johansen estimator - we found deviations up to a factor of 3.

Combining bias and variability for estimation of RR for AE probabilities, we analyzed the impact of using different estimators on making decisions about effect size based on the criteria put forward by IQWiG. We found that the number of AE types for which a given estimator deviates from the decision on effect size based on the gold-standard Aalen-Johansen estimator is non-negligible, and that discrepancies by more than one category also occur quite often. This is likely a consequence of both, the bias we see for estimation of RR for some of the estimators and the variability.

The analysis based on hazards reveals that the incidence density underestimates the RR for time to AE and time to a CE of death compared to the gold-standard Cox regression, while no obvious bias was observed for time to all CEs. The discrepancy in conclusions with regard to effect sizes drawn based on the ratio of incidence densities and the Cox regression HR appeared to be a bit less than for the estimators of RR based on AE probabilities. Still, the ratio of incidence densities cannot be considered a uniformly good approximation of the HR based on Cox regression.

Comparing the evidence categories derived from the two gold-standard
estimators, the Aalen-Johansen estimator of the RR and the Cox regression HR
for an AE we find quite some discrepancies. However, this is not
surprising, as the former is based on probability estimators and as such on cumulative measures integrating the two hazards relating to the primary event of interest (AE) and the potential CE, whereas the latter is an instantaneous measure only considering the AE hazard. In other words, this comparison reiterates the importance of accounting for CEs. There are now as many hazards as there are CEs, and outcome probabilities depend on all event-specific hazards. We are aware that in applications it might not be feasible to look at a hazard-based analysis for the AE and all CEs for every preferred term of AE, say. However, we recommend to consider such an analysis at least for AEs of special interest.

Finally, in an analysis of a composite endpoint with a single event of AE or CE, we find an overestimation of RR compared to the gold-standard one minus Kaplan-Meier estimator.

The organisational setup of SAVVY with developing program code centrally and running it within every sponsor organisation may be a template for collaborations with similar challenges. The setup avoided issues with data sharing and turned out to be highly efficient in bringing a large amount of data together.

Our empirical study does have shortcomings, some of which were to be anticipated in an opportunistic sample of randomized clinical trials. Inter alia, a large number of trials from oncology came with a high
  amount of CEs, which, in turn, led to comparable performances of arm comparisons based on incidence proportion and Aalen-Johansen. These shortcomings have been discussed in detail in Stegherr et al. \cite{onesample}, but this opportunistic ``real world'' sample allowed us
% . Using an opportunistic sample of
% randomized clinical trials from several sponsor companies, we have been able
% to illustrate possible consequences when quantifying the RR based on AE
% probabilities or hazards in a manner that ignores censoring or competing
% events. However, being opportunistic, the sample does not lend itself to
% straightforward generalizations. More than two thirds of the trials were from
% oncology. These came with a high amount of CEs, which, in turn,
% led to comparable performances of incidence proportion and Aalen-Johansen. The
% vast majority of AEs were classified as `common' or `very common', and AEs
% were also heterogeneous, coming from different therapeutic areas and were not
% necessarily treatment-related. These shortcomings were to be anticipated from
% an opportunistic sample, but it was our aim in this 'real-world' setting
to investigate and demonstrate which biases can occur in practice when
estimating a RR.  The observed results motivate future empirical
investigations on how to quantify RR with the aim of better
generalizability. As a further point, it was not the aim of this investigation
to accurately estimate RRs, but to compare different estimators. Our present
study does not allow for a meaningful comparison of results in different
diseases. Follow-up investigations concentrating on trials in specific disease
areas are planned.

Methodological restrictions include a focus on AE occurrence in a
  time-to-first-event setting, which does not consider recurrent AEs and often
  excludes AEs after treatment discontinuation. A more detailed discussion in
  Stegherr et al. \cite{onesample} stresses both the need to consider more
  complex event histories and the need to still account for CEs in such
  considerations.
% A methodological restriction is that we have focused our investigation on an
% analysis which mostly does not consider AEs after treatment discontinuation
% due to e.g. disease progression in oncology. This restriction is, in
% particular, due to trial design when treatment discontinuation leads to
% stopping AE recording after a pre-specified time period. In addition, in
% oncology, it is not uncommon that patients enter a different clinical trial
% after progression which further complicates matters. Another methodological
% restriction is that we did not consider recurrent AEs, but only first
% events. It is desirable to consider more complex event histories, also beyond
% time-to-first-event. However, any such consideration will need to account for
% CEs (and censoring), and our investigation therefore also informs
% methodological considerations for analysing such more complex event
% histories.
In other words, both AEs after treatment discontinuation and recurrent AEs will still be subject to CEs, and this must be accounted for when comparing arms.

We focused here on two effect measures, a comparison of probability estimates and the HR at different evaluation times. This, because we consider these two as the overwhelmingly used effect measures to quantify safety effects and we are interested in assessing potential biases of those often used methods, acknowledging that often in applications AE hazards might actually not be proportional. If non-proportional hazards for estimation of RRs for AE risk are a concern then we recommend to rely on comparison of AE probabilities at a fixed timepoint or of cumulative incidence functions.
%Alternative effect measures for time-to-event endpoints could have been considered as well, a potential candidate being restricted mean survival time (RMST). However, to the best of our knowledge, RMST is primarily used in the context of efficacy analyses and also has its challenges, e.g. interpretability and choice of the restriction timepoint, see the discussion in Freidlin and Korn\cite{freidlin_19}. Furthermore, RMST may be suited for a comprehensive composite endpoint, but extension to CEs is, although possible, not straightforward.

Replacing the often used incidence proportion by the gold-standard Aalen-Johansen estimator, while conceptually and empirically indicated, requires careful discussion in trial teams to define CEs, and a more granular data collection. In addition to the date of AEs one also needs to collect dates of CEs, which may lead to more missing data, e.g. unknown date of loss to follow-up.

In line with Stegherr et al. \cite{onesample}, our recommendation is to ``play
it safe'' when estimating RRs in a time-to-first-event analysis and neither
hope for a small amount nor a large amount of CEs nor a favorable interplay of
the distributions of the times of AEs, CEs, and censorings. In the former
case, one minus Kaplan-Meier might work well, while in the latter two cases
the incidence proportion might do so. However, in general the proportion of
CEs cannot exclusively explain how an estimator performs compared to the
gold-standard Aalen-Johansen estimator. Therefore, playing it safe, we
recommend using RR based on the Aalen-Johansen estimator for AE probabilities
and the HR from Cox regression for all types of events that are typically
considered in a time-to-first-event analysis. Guidelines for reporting AEs
should therefore advocate the Aalen-Johansen estimator instead of incidence
proportion, incidence density and one minus Kaplan-Meier, see Stegherr et
al. \cite{onesample} for an extended discussion on how to update
guidelines. {\color{black}We recommend that Aalen-Johansen estimators be used
  in product labels to quantify AE incidence.} A request for results from Cox regression
in guidelines should be complemented by also requesting results for
CE-specific hazards.

{\color{black}The choice of Aalen-Johansen over Kaplan-Meier has links to
  causal inference and debates on ``cause removal'' \cite{kalbfleisch2002}. In the common random
  censorship model, assuming independence of time-to-event and
  time-to-censoring, Kaplan-Meier can be given a causal interpretation under
  the intervention ``do(no censoring)'', i.e., Kaplan-Meier approximates the
  distribution of the data we would have seen in a world without censoring. It
  is therefore tempting to interpret Kaplan-Meier censoring a CE in a world
  where the CE is not active (i.e., cause removal). An important distinction is that the former
  intervention acts on our observational process, while the latter ``CE
  removal'' would affect the patient. Ignoring the question whether such an
  intervention of ``CE removal'' exists theoretically, the statistical question
  is whether Kaplan-Meier censoring a CE can be justified from a causal
  perspective. The recent work by Young et al.\cite{young2020causal} gives
  technical counterarguments in an estimand framework. The subject matter
  intuition is that an intervention on the observation process differs from an
  intervention affecting the patient.}

In this context, we also note as a limitation that our paper has only
  marginally addressed the current debate on estimands and, in that sense, is
  a more traditional statistical paper. A major reason behind this is the
  state of affairs in the statistical analyses of AEs. The ongoing debate on
  wether to use incidence proportions or incidence densities is far from a
  serious consideration of estimands. Our approach here has been that the
  incidence proportion implicitly defines the estimand, because it estimates
  $P(\mbox{AE in\ }[0,t])$ in situations not additionally complicated by
  varying follow-up times or censoring. Starting from this, the statistical
  aim of this paper has been to establish estimation and comparison of
  $P(\mbox{AE in\ }[0,t])$ in more complex data situations which are, in
  particular, characterized by the presence of competing events.

  To return to the possible interpretation of Kaplan-Meier under the
  intervention ``do(no censoring)'', we stress that, e.g., Kaplan-Meier
  estimation censoring CEs can not simply be aligned with a ``hypothetical
  estimand'' for the intercurrent CE. The statistical issue is that more
  complex causal methodology would be required. The conceptual issue is that
  hypothetical CE removal may be a mere hypothesis in one case and bear a real
  life interpretation in another case.

\section*{Conflict of interests}
KR and TK are employees of F. Hoffmann-La Roche (Basel, Switzerland). VJ and CDB are employees of Novartis Pharma AG (Basel, Switzerland). AA, AB, LE, KK, FL, MT, YZ are employees of Merck KGaA (Darmstadt, Germany), Bayer AG (Wuppertal, Germany), Janssen-Cilag GmbH (Neuss, Germany), Bristol-Myers-Squibb GmbH \& Co. KGaA (M\"unchen, Germany), Pfizer Deutschland (Berlin, Germany), Boehringer Ingelheim Pharma GmbH \& Co. KG (Ingelheim, Germany), Eli Lilly and Company (Indianapolis, USA), respectively. TF has received personal fees for consultancies (including data monitoring committees) from Bayer, Boehringer Ingelheim, Janssen, Novartis and Roche, all outside the submitted work. JB has received personal fees for consultancy from Pfizer, all outside the submitted work. CS has received personal fees for consultancies (including data monitoring committees) from Novartis and Roche, all outside the submitted work. The companies mentioned contributed data to the empirical study. RS has declared no conflict of interest.

\section*{Data Availability Statement}
Individual trial data analyses were run within the sponsor organisations using SAS and R software provided by the academic project group members. Only aggregated data necessary for meta-analyses were shared and meta-analyses were run centrally at the academic institutions. The aggregated data of the clinical trials are not publicly available due to confidentiality restrictions.

A markdown file providing exemplary code to compute all the estimators discussed in this paper for a given dataset is available on github: \url{https://github.com/numbersman77/AEprobs}. The corresponding output is available as html file: \url{https://numbersman77.github.io/AEprobs/SAVVY_AEprobs.html}.
%\section*{Author’s contributions}
%JB, CS and TF conceived the idea for the empirical study. RS was in charge of the analysis of the aggregated data and took part in interpretation and drafting of the manuscript together with CS, TF and JB. RS, CS, VJ, KR, FLe, JB and TF contributed to the design of the empirical study. CS, AB, LE, TK, KK, FLa, FLe, AL, CN and FV supervised the trial level analyses within the organisations. All authors critically reviewed the manuscript and approved its final version.

\bibliographystyle{SageV}
\bibliography{stat}

\def\cprime{$'$}
\begin{thebibliography}{10}
\providecommand{\url}[1]{\texttt{#1}}
\providecommand{\urlprefix}{URL }
\expandafter\ifx\csname urlstyle\endcsname\relax
  \providecommand{\doi}[1]{DOI:\discretionary{}{}{}#1}\else
  \providecommand{\doi}{DOI:\discretionary{}{}{}\begingroup
  \urlstyle{rm}\Url}\fi
\providecommand{\eprint}[2][]{\url{#2}}

\bibitem{proctor_16}
Proctor T and Schumacher M.
\newblock Analysing adverse events by time-to-event models: the cleopatra
  study.
\newblock \emph{Pharm Stat} 2016; 15(4): 306--314.
\newblock \doi{10.1002/pst.1758}.
\newblock \urlprefix\url{http://dx.doi.org/10.1002/pst.1758}.

\bibitem{koller_12}
Koller MT, Raatz H, Steyerberg EW et~al.
\newblock Competing risks and the clinical community: irrelevance or ignorance?
\newblock \emph{Stat Med} 2012; 31: 1089--1097.
\newblock \doi{10.1002/sim.4384}.

\bibitem{onesample}
Stegherr R, Schmoor C, Beyersmann J et~al.
\newblock {Survival analysis for AdVerse events with VarYing follow-up times
  (SAVVY) -- estimation of adverse event risks}.
\newblock \emph{Trials} 2021; 22(420).
\newblock \doi{https://doi.org/10.1186/s13063-021-05354-x}.

\bibitem{allignol_16}
Allignol A, Beyersmann J and Schmoor C.
\newblock Statistical issues in the analysis of adverse events in time-to-event
  data.
\newblock \emph{Pharm Stat} 2016; 15: 297--305.
\newblock \doi{10.1002/pst.1739}.

\bibitem{schuster_20}
Schuster NA, Hoogendijk EO, Kok AAL et~al.
\newblock Ignoring competing events in the analysis of survival data may lead
  to biased results: a nonmathematical illustration of competing risk analysis.
\newblock \emph{J Clin Epidemiol} 2020; 122: 42--48.

\bibitem{stegherr_20}
Stegherr R, Schmoor C, L\"ubbert M et~al.
\newblock {{E}stimating and comparing adverse event probabilities in the
  presence of varying follow-up times and competing events}.
\newblock \emph{Pharm Stat} 2021; 20(6): 1125--1146.

\bibitem{unkel_19}
Unkel S, Amiri M, Benda N et~al.
\newblock On estimands and the analysis of adverse events in the presence of
  varying follow-up times within the benefit assessment of therapies.
\newblock \emph{Pharm Stat} 2019; 18: 166--183.
\newblock \doi{10.1002/pst.1915}.

\bibitem{stegherr_19}
Stegherr R, Beyersmann J, Jehl V et~al.
\newblock Survival analysis for adverse events with varying follow-up times
  (savvy): Rationale and statistical concept of a meta-analytic study.
\newblock \emph{Biom J} 2021; 63: 650--670.
\newblock \doi{10.1002/bimj.201900347}.

\bibitem{budin_15}
Budin-Ljosne I, Burton P, Isaeva J et~al.
\newblock {{D}ata{S}{H}{I}{E}{L}{D}: an ethically robust solution to
  multiple-site individual-level data analysis}.
\newblock \emph{Public Health Genomics} 2015; 18(2): 87--96.

\bibitem{austin_17a}
Austin PC and Fine JP.
\newblock Accounting for competing risks in randomized controlled trials: a
  review and recommendations for improvement.
\newblock \emph{Stat Med} 2017; 36: 1203--1209.
\newblock \doi{10.1002/sim.7215}.

\bibitem{Gool:Leis:Crow:Stor:esti:1999}
Gooley TA, Leisenring W, Crowley J et~al.
\newblock Estimation of failure probabilities in the presence of competing
  risks: {N}ew representations of old estimators.
\newblock \emph{Stat Med} 1999; 18: 695--706.

\bibitem{pocock_02}
Pocock SJ, Clayton TC and Altman DG.
\newblock {{S}urvival plots of time-to-event outcomes in clinical trials: good
  practice and pitfalls}.
\newblock \emph{Lancet} 2002; 359(9318): 1686--1689.

\bibitem{Terry}
Therneau T and Grambsch P.
\newblock \emph{{Modeling Survival Data: Extending the Cox Model.}}
\newblock {Springer, New York}, 2000.

\bibitem{latouche_13}
Latouche A, Allignol A, Beyersmann J et~al.
\newblock A competing risks analysis should report results on all
  cause-specific hazards and cumulative incidence functions.
\newblock \emph{J Clin Epidemiol} 2013; 66(6): 648--653.

\bibitem{iqwigmethodenpaper}
{IQWiG}.
\newblock {General Methods, Version 5.0. Institute of Quality and Efficiency in
  Health Care}, 2017.
\newblock
  \urlprefix\url{https://www.iqwig.de/en/methods/methods-paper.3020.html}.
\newblock Https://www.iqwig.de/en/methods/methods-paper.3020.html.

\bibitem{schemper_96}
Schemper M and Smith TL.
\newblock A note on quantifying follow-up in studies of failure time.
\newblock \emph{Control Clin Trials} 1996; 17: 343--346.

\bibitem{clark_02}
Clark TG, Altman DG and De~Stavola BL.
\newblock Quantification of the completeness of follow-up.
\newblock \emph{Lancet} 2002; 359: 1309--1310.

\bibitem{varadhan_10}
Varadhan R, Weiss CO, Segal JB et~al.
\newblock Evaluating health outcomes in the presence of competing risks: a
  review of statistical methods and clinical applications.
\newblock \emph{Med care} 2010; 48: S96--105.
\newblock \doi{10.1097/MLR.0b013e3181d99107}.

\bibitem{schumacher2016competing}
Schumacher M, Ohneberg K and Beyersmann J.
\newblock Competing risk bias was common in a prominent medical journal.
\newblock \emph{J Clin Epidemiol} 2016; 80: 135--136.

\bibitem{phillips_20}
Phillips R and Cornelius V.
\newblock Understanding current practice, identifying barriers and exploring
  priorities for adverse event analysis in randomised controlled trials: an
  online, cross-sectional survey of statisticians from academia and industry.
\newblock \emph{BMJ Open} 2020; 10(6).

\bibitem{kalbfleisch2002}
Kalbfleisch JD and Prentice RL.
\newblock \emph{The statistical analysis of failure time data}.
\newblock John Wiley \& Sons, Hoboken, New Jersy, 2002.

\bibitem{young2020causal}
Young JG, Stensrud MJ, Tchetgen~Tchetgen EJ et~al.
\newblock A causal framework for classical statistical estimands in
  failure-time settings with competing events.
\newblock \emph{Stat Med} 2020; 39(8): 1199--1236.

\end{thebibliography}

\clearpage

\end{document}